\documentclass{ar-1col}
\usepackage[numbers]{natbib}

\usepackage{amsfonts} 
\usepackage{amssymb} 
\usepackage{amsmath} 
\usepackage{verbatim} 
\usepackage[legalpaper,margin=1.5in]{geometry}
\usepackage{hyperref}

\newcommand{\gsim}{\mathrel{\rlap{\lower4pt\hbox{\hskip1pt$\sim$}}\raise1pt\hbox{$>$}}}
\newcommand{\lsim}{\mathrel{\rlap{\lower4pt\hbox{\hskip1pt$\sim$}}\raise1pt\hbox{$<$}}}

\jname{Annu. Rev. Nucl. Part. Sci}
\jvol{70}
\jyear{2020}
\doi{10.1146/annurev-nucl-101918-023530}
\firstpagenote{When citing this paper, please use the following: Hoang AH. 2020. What is the top quark mass? \textit{Annu. Rev. Nucl. Part. Sci.} 70: Submitted. DOI:10.1146/annurev-nucl-101918-023530}

\begin{document}

\markboth{Andr\'e H. Hoang}{What is the Top Quark Mass?}

\title{What is the \\Top Quark Mass?}

\author{Andr\'e H. Hoang$^{1,2}$
\affil{$^1$University of Vienna\\ Faculty of Physics, Boltzmanngasse 5,  A-1090 Wien, Austria\\email: andre.hoang@univie.ac.at}
\affil{$^2$Erwin Schr\"odinger International Institute for Mathematical Physics\\
	Boltzmanngasse 9, A-1090 Wien, Austria\\[5mm]UWThPh-2020-1}}

\begin{abstract}
In this review I give an overview on the conceptual issues involved in the question how to interpret so-called `direct top quark mass measurements', which are based on the kinematic reconstruction of top quark decay products at the Large Hadron Collider (LHC). 
These measurements quote the top mass parameter $m_t^{\rm MC}$ of Monte-Carlo event generators with current uncertainties of around $0.5$\,GeV.  
At present time the problem of finding a rigorous relation between $m_t^{\rm MC}$ and top mass renormalization schemes defined in field theory is unresolved and touches perturbative as well as nonperturbative aspects and the limitations of state-of-the-art Monte-Carlo event generators.
I review the status of LHC top mass measurements, illustrate how conceptual limitations enter and 
explain a controversy that has permeated the community in the context of the interpretation problem related to $m_t^{\rm MC}$.
Recent advances in acquiring first principle insights are summarized, and it is outlined what else has to be understood to fully resolve the issue. For the time being, I give a recommendation how to deal with the interpretation problem when making top mass dependent theoretical predictions.
\end{abstract}

\maketitle

\tableofcontents

\section{INTRODUCTION}
\label{sec:intro}

The top quark is the heaviest particle of the Standard Model of elementary particle physics (SM).
The currently most precise determinations of its mass come from so-called `direct measurements'. These are based on the experimental kinematic reconstruction of the final-state top quark decay products (which are bottom quark jets, light quark jets from $W$ boson decays and leptons) and the comparison of kinematic distributions one can construct from the 4-momenta of the decay products with descriptions of the same quantities obtained from multipurpose Monte Carlo event generators (MMC). These measurements determine the top mass parameter of the MMC and yield a world average  of $m_t^{\rm MC} = 172.9 \pm 0.4$\,GeV~\cite{Tanabashi:2018oca}. This amounts to an impressive relative precision of $0.2\%$ and makes the top quark mass the most precisely known parameter in the strong interaction sector of the SM, called quantum chromodynamics (QCD). For the high luminosity phase of the LHC (HL-LHC) it is projected that uncertainties as small as $200$\,MeV can be reached from direct top mass measurements~\cite{Azzi:2019yne}. 

The major portion of the top quark's mass is generated through the electroweak Higgs mechanism~\cite{Higgs:1964pj,Englert:1964et,Guralnik:1964eu} which also gives all other elementary particles of the SM their masses. The precise knowledge of the elementary particle masses and their couplings is an important element in consistency tests of the SM and in indirect searches of physics beyond the SM. Because the hopes for discoveries of non-SM elementary particles at the LHC have up to now not been fulfilled, indirect new physics searches, which focus on finding deviations between experimental data and SM predictions, have become increasingly important. This requires a high level of precision and a thorough and systematic understanding of subtle experimental as well as theoretical issues. In this context the top quark plays a special role because its large mass makes it a highly sensitive probe of the structure of the SM Higgs sector and an important ingredient in models of physics beyond the SM. In this context it is the electroweak part of the top quark mass one seeks to know with the highest possible precision. It is frequently stated that, due to its small lifetime ($1/\tau_t=\Gamma_t= 1.42^{+0.19}_{-0.15}$~GeV~\cite{Tanabashi:2018oca}) the top quark, even though it has strong color charge, is protected from low-scale hadronization effects, approximately behaving as a free particle. 

The results for $m_t^{\rm MC}$ obtained from the direct measurements were frequently identified with the so-called top quark pole mass $m_t^{\rm pole}$, which is a popular renormalization scheme used for perturbative QCD computations at next-to-leading order (NLO) and beyond. The pole mass encodes, strictly within perturbation theory, the notion of the kinematic rest mass of the top quark as a real on-shell particle. The identification appeared natural because the top mass sensitivity of the kinematic distributions entering the direct measurement analyses is coming from resonance and endpoint structures that can be seen to be related to the kinematic properties of a top quark particle with a definite mass. I refer to distributions of this kind as `observables with kinematic (top) mass sensitivity'.
With the identification of $m_t^{\rm MC}$ and $m_t^{\rm pole}$ precise higher-order predictions for the SM electroweak potential~\cite{Cabibbo:1979ay,Alekhin:2012py,Buttazzo:2013uya,Andreassen:2014gha,Branchina:2013jra} have been made. These, together with precise measurements of the Higgs boson mass~\cite{Tanabashi:2018oca}, indicated that the SM is in a metastable state.\footnote{The coupling of the top quark to the Higgs boson generates the large electroweak portion of the mass of the top quark. The top quark mass conversely causes large quantum corrections to the Higgs self-coupling which determines the Higgs mass and also the stability properties of the potential for the Higgs boson field and the SM vacuum. These quantum effects decrease the Higgs self coupling for a larger top mass with the possibility to destabilize the vacuum~\cite{Isidori:2001bm}.} 
However, in recent years a discussion emerged whether, considering a precision of $0.5$~GeV or better, the available NLO (and higher order) perturbative calculations and NLO-matched MMCs indeed control the QCD dynamics affecting the top quark mass and the way the direct measurement observables depend on it
sufficiently well, to justify the identification of $m_t^{\rm MC}$ with the pole mass~\cite{Hoang:2008xm,Moch:2010rh,Deliot:2010ey,Alekhin:2012py,Juste:2013dsa,Moch:2014tta,Corcella:2014rya,Hoang:2014oea,Weinzierl:2015gua,Boos:2015bta,delDuca:2015gca,Corcella:2015kth,Azzi:2017iih}.
Here, the direct measurements, being the most precise and known to rely essentially entirely on the parton-shower and hadronization dynamics of the MMCs, were discussed most intensely. 
I call the associated set of physical issues the {\bf 'top mass interpretation problem'}.
The top mass interpretation problem is the question of the precise relation between $m_t^{\rm MC}$ and more fundamental and field theoretic mass definitions such as the pole mass, the $\overline{\mbox{MS}}$ mass or other mass schemes. The origin of the problem is that the simple picture of a free top quark, that directly governs the visible structures in distributions with kinematic top mass sensitivity, is too naive and that the effects of QCD and electroweak quantum fluctuations must be accounted for to high precision. These quantum effects are governed by low energy scales at the level of 1~GeV even though the top mass is extremely large, and they can directly affect the extracted top mass if not described theoretically in an adequate way. What makes matters subtle is that the low-energy QCD dynamics is difficult to control theoretically because of large higher-order perturbative corrections and nonperturbative effects. The top mass interpretation problem emerges because the top mass sensitive kinematic distributions used for the direct measurements are so complicated that with the current technology their theoretical description can only be provided by MMCs. In the current generation of MMCs, however, the
theoretical precision and quality of the low-energy parton-shower and hadronization dynamics 
cannot yet be systematically controlled at a level such that the identification of the top mass parameter $m_t^{\rm MC}$ with a
field theoretic mass scheme such as the pole mass can be proven from first principles.

Probably the most confusing aspect of the emerging discussions has been that no consensus has been reached on how to estimate and even formulate the uncertainty associated to the top mass interpretation problem and how to deal with it in practical applications, see e.g.\ Ref.~\cite{Azzi:2019yne}. 
Furthermore, the issue has not been discussed in a coherent fashion in the community and the advocated points of view were slightly shifting over time. I call this aspect of the interpretation problem {\bf 'the controversy'} because it is related to different views on the relevance of the physical aspects of the interpretation problem, but does not contribute in any way to a resolution of the physical questions.  
Meanwhile a number of alternative top mass measurement methods were devised for the LHC, partly with the motivation of applying methods that are not or in a different way affected by the interpretation problem. These methods are still less precise than the direct measurements and have their own subtleties once their precision increases.
The situation is reflected in an interesting way in the Review of Particle Physics~\cite{Tanabashi:2018oca}, where the top is the only quark for which three different masses are quoted in the particle listings.
I want to emphasize clearly, however, that in hadron-hadron collisions, where the underlying hard interactions that are
the basis of the observables unavoidably involve partons in non-singlet color configurations, the conceptual issues that affect the direct measurements are eventually emerging for all top mass measurements methods once a precision of $0.5$~GeV or better is reached.

The interpretation problem consists of a complex set of issues and requires significant theoretical progress on multiple fronts, predominantly beyond the realm of fixed-order perturbative calculations. The issue can be resolved in a straightforward and fully transparent way only once at least next-to-leading-logarithmic (NLL) precise parton showers and MMC event generators have become available (for the observables entering the top mass measurements). The latter should be capable of describing the top quark decay and the non-perturbative aspects of color neutralization in a systematic manner such that the field theory aspects of the MMC top mass parameter (and even the strong coupling parameter) are well-defined and can be determined from a simple computation. Such developments clearly require a dedicated and long-term effort that will also benefit all other aspects of collider physics. Even if this may be a bit too much to hope for, I believe that, by addressing these issues through dedicated studies, already much can be learned, so that the controversial situation can be lifted at least partially for some of the direct top mass measurement methods.

In this review, I explain the questions involved in the top mass interpretation problem from a physical and conceptual perspective. I hope that the review allows the reader to gain a better understanding of the physical and systematic aspects of the interpretation problem (and also the controversy) and an appreciation of the problems to be resolved. 
I am trying to be as non-technical as possible, but the problem is of subtle theoretical nature.
For simplicity all formulas shown are either understood generic or truncated at ${\cal O}(\alpha_s)$ or NLO, even though higher order corrections are mostly known. All numerical results quoted have been computed including all available higher order corrections and I use $\alpha_s^{(n_\ell=5)}(M_Z)=0.118$ as the reference input for the strong coupling. I apologize for any missing 
references.

The review is organized as follows.
In Sec.~\ref{sec:mass} the physical aspects of different top quark renormalization schemes are reviewed. 
This section serves as a basis of the following discussions. But 
I emphasize that a mere discussion on top mass schemes does not resolve the top mass interpretation problem. 
In Sec.~\ref{sec:status} an overview is given on the current status of top quark mass measurements and the  theoretical tools employed. Here I focus on the limitations of the theoretical tools, which are the origin of the top mass interpretation problem, and the role they play for the other top mass measurement methods.  
In Sec.~\ref{sec:controversy} I then phrase the controversy in a set of formulae which can be discussed in a concrete way.  
In Sec.~\ref{sec:recent} recent work is reviewed which quantifies the interpretation problem numerically. In particular, I discuss the recent work of Ref.~\cite{Hoang:2018zrp} where some first conceptual and concrete analytical insights into the perturbative (parton level) aspects of the interpretation problem were gained. 
Finally, in Sec.~\ref{sec:conclusion} I wrap up and conclude with a recommendation on how to practically deal with the top mass interpretation problem at this time. Appendix~\ref{sec:AppA} contains a basic glossary.

\section{MASS EXTRACTION AND RENORMALIZATION SCHEMES}
\label{sec:mass}

\subsection{The Principle of Top Mass Determinations}
\label{sec:principle}

Since the top quark - just as all other quarks which carry strong interaction color charge - is not a physically observable particle, its mass is a quantity that needs to be defined from a theoretical prescription in quantum field theory, called a renormalization scheme or just a mass scheme. The choice of a scheme is in principle arbitrary. The usefulness and systematics of this concept arises from the facts that we can make precise predictions for a physical observable in a given scheme and that the predictions in
different schemes can be related to each other by a theoretical calculation. Since contemporary high-energy physics for the most part considers observables where perturbative computations can be used, usually only quark mass schemes defined strictly within perturbation theory are considered. This is the view commonly accepted in high-energy collider physics, and I also adopt it in this review.
So given two top mass schemes (called $m_t^A$ and $m_t^B$), the relation between the two can be described by a perturbative series of the form
\begin{equation}
\label{eq:ABconversion}
m_t^A - m_t^B \, = \, \sum_{n=1} c_n \alpha_s^n(\mu)\,.
\end{equation}
For simplicity I only indicate powers of the strong coupling $\alpha_s$ and suppress contributions from the electromagnetic and weak couplings. The precision of the relation is limited by the ability to calculate and then to evaluate the truncated series in a meaningful way. I emphasize this because perturbative series in non-Abelian gauge theories such as QCD are in general asymptotic and not convergent. I will come back to this point below.
Given a top mass sensitive observable $\sigma$, where I mostly refer to different kinds of cross sections, the corresponding perturbative (and also asymptotic) series, called `parton level' cross section, can be written as $\hat\sigma(Q,m_t^X,\alpha_s(\mu),\mu;\delta m^X)$. The energy  $Q$ of the process, the top mass $m_t^X$ in scheme $X$, the strong coupling $\alpha_s(\mu)$ and its renormalization scheme $\mu$ appear as explicit arguments. Furthermore, the separated argument $\delta m^X$ stands for the dependence of the series on the scheme choice $X$. By construction, the perturbative series 
in the two mass schemes are formally equivalent:\footnote{I suppress the dependence on the masses of other quarks or leptons}
\begin{equation}
\hat\sigma(Q,m_t^A,\alpha_s(\mu),\mu;\delta m^A) \, = \, 
\hat\sigma(Q,m_t^B,\alpha_s(\mu),\mu;\delta m^B)\,.
\end{equation}
However, in practice they differ due to the truncation of the series and our limited ability to calculate and sum the series. 

For the strong coupling $\alpha_s$ the freedom of scheme exists as well. Within the commonly accepted paradigm of using dimensional regularization and the so-called $\overline{\rm MS}$ prescription for $\alpha_s$ (which are explained in the next subsection), this is signified by the dependence on the renormalization scale $\mu$. A useful aspect of the coupling $\alpha_s(\mu)$ is, that one can interpret $\mu$ as the momentum scale above which all quantum corrections to the fundamental QCD gluon interactions are absorbed into $\alpha_s(\mu)$. So, frequently, for the choice $\mu\sim Q$ (particularly when $Q\gg m_t$) the resulting perturbation series behave quite well. This way also an important set of logarithmic corrections is summed up to all orders in perturbation theory. The numerical differences between different considered reasonable scheme choices are then typically used to estimate the theoretical uncertainties of the parton level cross section.

For the extraction of the top mass (and any other QCD parameter) from an experimentally measured cross section $\sigma^{\rm exp}$, however, the parton level cross section does not provide the full answer and one has to also account for nonperturbative corrections:
\begin{equation}
\label{eq:sigma}
\sigma^{\rm exp} = \hat\sigma(Q,m_t^X,\alpha_s(\mu),\mu;\delta m^X) \, + \, 
\sigma^{\rm NP}(Q,\Lambda_{\rm QCD})\,.
\end{equation}
Here, $\Lambda_{\rm QCD}$ stands for a nonperturbative scale with typical size of a few hundred MeV that governs the magnitude of the nonperturbative correction $\sigma^{\rm NP}$. The form of Eq.~(\ref{eq:sigma}) is schematic and also accounts for the nonperturbative effects in the parton distribution functions needed to calculate cross sections at the LHC. For most cross sections one has $Q\gg\Lambda_{\rm QCD}$ and typically $\hat\sigma > \sigma^{\rm NP}$. 
Since we consider mass schemes strictly defined within perturbation theory, the relations between mass schemes shown in  Eq.~(\ref{eq:ABconversion}) are free of nonperturbative corrections, so that switching between mass schemes does never modify the structure and the content of $\sigma^{\rm NP}$.
The precision for the top mass extraction depends on the ability to calculate the perturbative cross section $\hat\sigma$ and to determine the nonperturbative correction $\sigma^{\rm NP}$. Likewise, the uncertainty in the extracted top mass arises from the combined uncertainties in $\hat\sigma$ and $\sigma^{\rm NP}$, where one has to keep in mind that $\sigma^{\rm NP}$ is per se not responsible for the top mass dependence of the observable $\sigma^{\rm exp}$.
The most preferred observables for top mass measurements (and in general), are those where $\sigma^{\rm NP}$ vanishes as $(\Lambda_{\rm QCD}/Q)^{n}$, with $n\ge 2$ for the limit $\Lambda_{\rm QCD}/Q\to 0$ because then the contributions of 
$\sigma^{\rm NP}$ can be very small (because $Q,m_t\gg \Lambda_{\rm QCD}$).
An example of such a ``clean'' observable is the total inclusive $t\bar t$ cross section in $e^+e^-$ annihilation for which the {\it observable-initiating hard reaction} is the production of a color-singlet $t\bar t$ pair via the process $e^+e^-\to \gamma,Z\to t\bar t$.\footnote{The inclusive $t\bar t$ cross section in $e^+e^-$ annihilation at c.m.\ energies $Q$ close to production threshold, $Q\approx 2m_t$, constitutes the most precise top mass measurement method at a future $e^+e^-$ collider. Theoretical cross section predictions~\cite{Fadin:1988fn,Strassler:1990nw,Jezabek:1992np,Hoang:2000yr,Hoang:2013uda,Beneke:2015kwa} have reached uncertainties at the level of several percent and allow for top mass determinations with uncertainties at the level of $50$~MeV or better~\cite{Seidel:2013sqa,Horiguchi:2013wra,Vos:2016til,Abramowicz:2018rjq}. Because the production of $t\bar t$ pairs in color-octet configurations is strongly suppressed, the effects of soft QCD radiation are strongly suppressed as well. This also applies to the $t\bar t\gamma$ final state analyzed in Ref.~\cite{Boronat:2019cgt}.} Here, $n=4$ and the nonperturbative corrections are negligible for most applications.
Unfortunately, at the LHC such clean cross sections do not exist. This is because non-singlet color configurations are unavoidable for the observable-initiating hard reactions when partons (i.e.\ quarks and gluons that emerge from the colliding protons) appear in the scattering initial state and when jet formation is crucial for the construction of an observable. Therefore, color neutralization processes that are linearly sensitive to soft and nonperturbative momenta are unavoidable, and $\sigma^{\rm NP}$ always depends linearly on $\Lambda_{\rm QCD}$. Thus we cannot get around dealing with $\sigma^{\rm NP}$ at the LHC.

\vspace{8mm}

\begin{figure}[h]
	\includegraphics[scale=1]{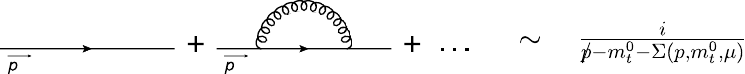}
	\vskip 3mm
	\caption{\label{fig:selfenergy}Self-energy at NLO for the top quark with four-momentum $p^\mu$. }
\end{figure}

\subsection{Top Mass Renormalization Schemes}
\label{sec:schemes}

In analogy to adopting an adequate choice for the renormalization scale $\mu$ of the strong coupling $\alpha_s(\mu)$, one also adopts an adequate top quark mass scheme. The central formal aspect of top quark mass renormalization is to absorb the UV divergence that arises in the NLO self-energy diagram, see the generic illustration in Fig.~\ref{fig:selfenergy}. Here
\begin{equation}
\label{eq:sigmaself}
\Sigma(p,m_t^0,\mu) \sim m_t^0\,\Big(\frac{\alpha_s(\mu)}{\pi}\Big)\left[\frac{1}{\epsilon} +\ln(4\pi e^{-\gamma_E})   
+ A^{\rm fin}(m_t^0/\mu)  \right] \,+\, \ldots
\end{equation}
displays the dominant contribution in the resonance limit $p^2\to m_t^2$ within a calculation in $d=4-2\epsilon$ space-time dimensions. Using $d$ dimensions is the standard way to regularize ultra-violet (UV) divergences in perturbative QCD computations and called `dimensional regularization'. The ellipsis stands for higher order corrections proportional to higher powers of the strong coupling $\alpha_s$, which are known to ${\cal O}(\alpha_s^4)$ from   Refs.~\cite{Tarrach:1980up,Gray:1990yh,Chetyrkin:1999ys,Chetyrkin:1999qi,Melnikov:2000qh,Marquard:2007uj}.
The term that is divergent in the limit $\epsilon\to 0$, which quantifies the UV divergence to be renormalized, and the finite term $A^{\rm fin}(m_t^0/\mu)$, are shown separately, and $m_t^0$ stands for the bare unrenormalized mass. In this context the term $A^{\rm fin}$, even though it is finite, contains a contribution from self-energy quantum fluctuations arising from soft (i.e.\ small) momenta in a {\it top resonance frame}.\footnote{
This is a reference frame where a top quark state within its finite-lifetime Breit-Wigner resonance region is having a non-relativistic average velocity. Such frames are frequently collectively called 'the top quark rest frame', but I will not adopt this jargon here, because it is not appropriate when discussing uncertainties in mass determinations much smaller than the top quark width $\Gamma_t(t\to bW)=1.4$~GeV.} These soft quantum corrections, which I will refer to as `ultracollinear' corrections in Sec.~\ref{sec:conceptual}, affect $m_t^0 A^{\rm fin}(m_t^0/\mu)$ linearly~\cite{Neubert:1993mb,Manohar:2000dt}. For example, giving the gluon a small test mass $\lambda$, which cuts off these soft momenta, one obtains~\cite{Ball:1995ni} 
\begin{equation}
\label{eq:Agluonmass}
\left.m_t^0 A^{\rm fin}(m_t^0/\mu)\right|_{{\rm gluon mass}\,\lambda} \,= \, \left.m_t^0 A^{\rm fin}(m_t^0/\mu)\right|_{\lambda=0} + \frac{2}{3}\lambda\, + \, {\cal O}(\lambda^2/m_t^0)\,.
\end{equation}
This is important to remember for the following, because perturbation theory does not work well in this regime.

\begin{figure}[t]
	\includegraphics[scale=0.3]{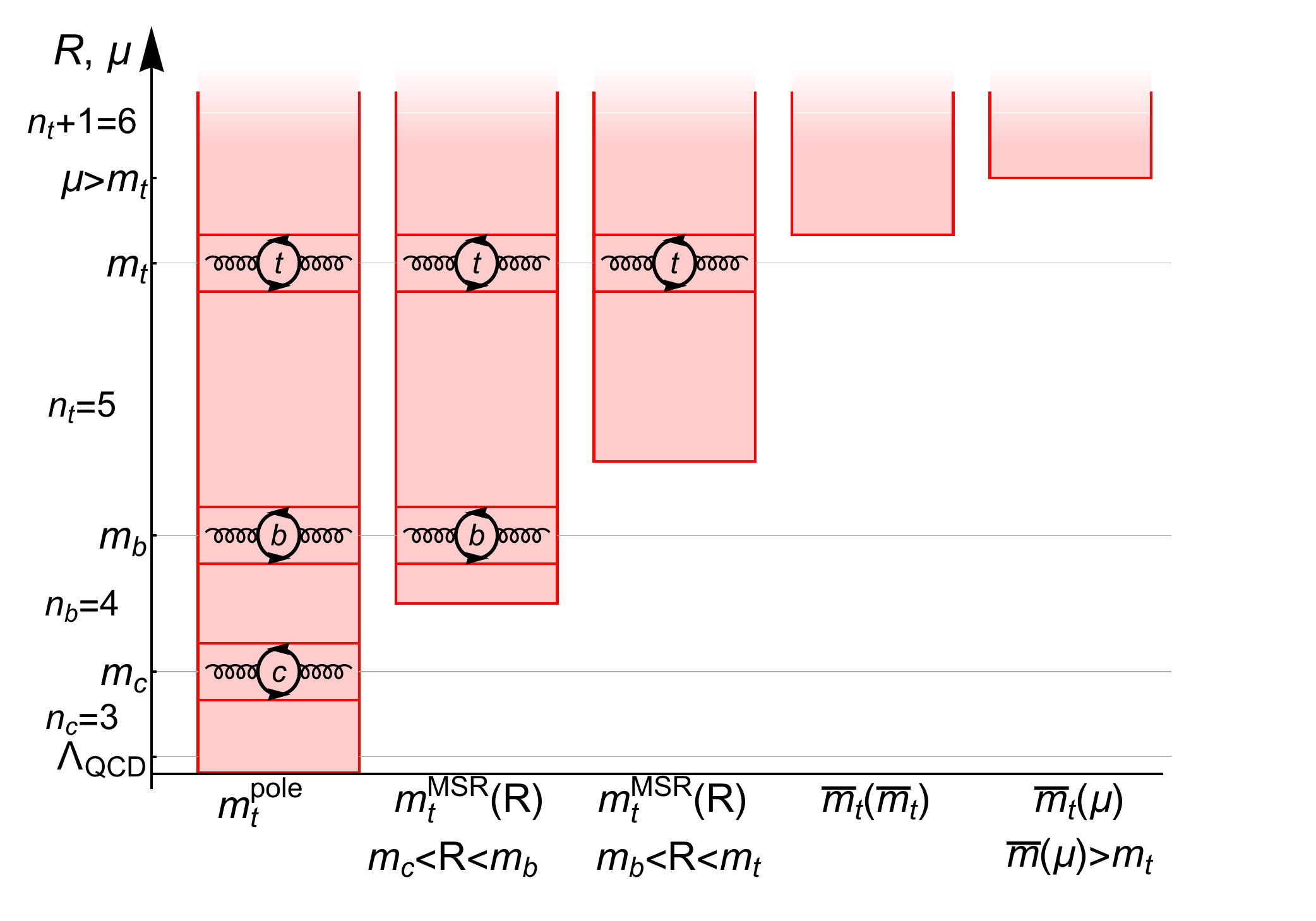}
	\caption{\label{fig:schemes1}The red shaded regions represent the self-energy contributions absorbed into the pole mass $m_t^{\rm pole}$, MSR mass $m_t^{\rm MSR}(R)$ and $\overline{\rm MS}$ mass $\overline m(\mu)$. The pole mass absorbs all contributions down to vanishing momenta, while MSR and $\overline{\rm MS}$ masses absorb only contributions above scales $R$ and $\mu$, respectively. From Ref.~\cite{Hoang:2017btd}.}
\end{figure}

The mass scheme that is closest to the concept of the strong coupling $\alpha_s(\mu)$ is the 
{\bf $\overline{\mbox{MS}}$ scheme} $\overline m_t(\mu)$. Here only the pure UV $1/\epsilon$ term (including the conventional term $\ln(4\pi e^{-\gamma_E})$) is absorbed into the renormalized mass, 
\begin{equation}
\overline{m}_t(\mu) = m_t^0 \left\{1+\Big(\frac{\alpha_s(\mu)}{\pi}\Big)\left[\frac{1}{\epsilon} +\ln(4\pi e^{-\gamma_E})  \right]\right\} \,+\, \ldots\,.
\end{equation}
The $\overline{\mbox{MS}}$ mass is $\mu$-dependent like $\alpha_s(\mu)$ and satisfies the renormalization group equation
\begin{equation}
\label{eq:muevolv}
\frac{\rm d}{{\rm d} \ln \mu} \,\overline{m}_t(\mu)\, = \, - \overline{m}_t(\mu) \,\Big(\frac{\alpha_s(\mu)}{\pi}\Big) \,+\, \ldots
\,,
\end{equation}
implying that $\overline m_t(\mu)$ depends logarithmically on $\mu$.
In analogy to $\alpha_s(\mu)$ we can interpret $\mu$ as the momentum scale (in a top resonance frame) above which all self-energy quantum corrections are absorbed into $\overline m_t(\mu)$, see Fig.~\ref{fig:schemes1} for a graphical illustration. The $\overline{\mbox{MS}}$ mass is therefore not affected by low-energy or nonperturbative quantum fluctuations and called a `short-distance mass'. 
The term  $A^{\rm fin}$ in Eq.~(\ref{eq:sigmaself}), which is not absorbed into the renormalized mass, still appears in the perturbative calculations of the process, and its 'bad' linearly sensitive small momentum contributions are known to cancel with other virtual (non-self-energy) corrections that are soft in a top resonance frame. This can be explicitly checked for any parton level cross section for a physical process involving top quarks considering the sum of all linear gluon mass terms as that displayed for $A^{\rm fin}$ in Eq.~(\ref{eq:Agluonmass}) coming from radiation that is soft in a top resonance frame.   
Setting $\mu$ to the physical scale of the process governing the mass dependence of an observable, together with a proper scale setting for the strong coupling, frequently yields good behavior for the perturbation series of $\hat\sigma$. 
The interpretation of $\overline m_t(\mu)$ mentioned above, however, only applies for observables where $\mu \gsim m_t$, which includes e.g.\ total inclusive cross sections at very high energies $Q\gg m_t$ or when the top effects are virtual such as for
the SM Higgs potential~\cite{Cabibbo:1979ay,Alekhin:2012py,Buttazzo:2013uya,Andreassen:2014gha,Branchina:2013jra}, the electroweak precision observables~\cite{Tanabashi:2018oca,Baak:2014ora} or the properties of $B$ mesons~\cite{Buras:1993wr}. Such observables can have a strong indirect top mass sensitivity, but not a kinematic mass sensitivity.

As already pointed out in Sec.~\ref{sec:intro}, observables with kinematic mass sensitivity are related to distributions of variables that show sharp threshold patterns such as resonances, shoulders or endpoints. Even though these patterns are initiated by hard reactions involving the large scales $m_t$ or $Q$, the observable mass dependence is in addition modified by dynamical QCD and electroweak quantum effects. The momentum scales governing these quantum effects are, however, soft, i.e.\ much smaller than $m_t$. I refer to this momentum scales generically as the scale `$R$'. So for observables with kinematic mass sensitivity we typically have $R\ll m_t$. 
The prototypical example is the invariant mass of jets coming from the hadronic decay of a top quark. Here, the scale $R$ governing the soft quantum effects I have been talking about is set by the width of the resonance visible in the invariant mass distribution. It is bounded from below by the top width $\Gamma_t$ or the experimental resolution. 
The left panel of Fig.~\ref{fig:ATLAStemplate} shows the top mass dependence of the reconstructed top invariant mass $m_t^{\rm reco}$ from LHC simulations carried out in Ref.~\cite{Aaboud:2018zbu}. From the distribution we can see that here we have $R\sim 30$~GeV. The high mass sensitivity arises from the location of the resonance structure. While the basic location and existence of the resonance is tied to the top quark mass, which is large, the details of the resonance shape, its width and the exact location is governed in addition by low-energy QCD and electroweak effects. Observables of this kind are the basis of the direct top mass measurements.

\begin{figure}[t]
	\includegraphics[scale=0.8]{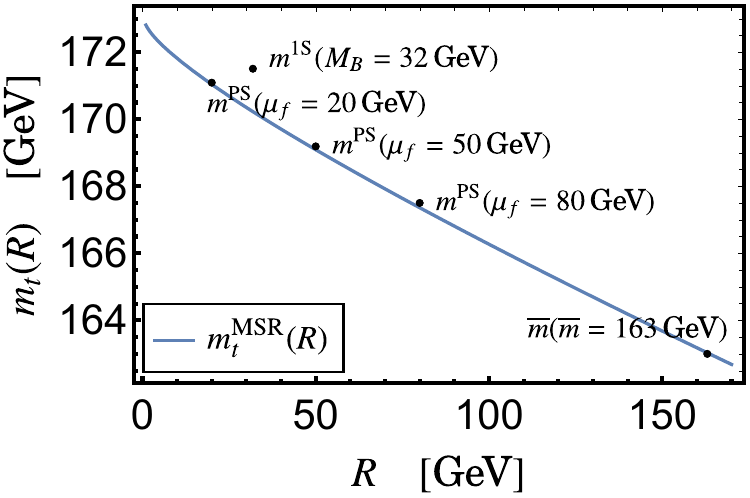}
	\caption{\label{fig:schemes2} MSR mass $m_t^{\rm MSR}(R)$ over $R$  (blue line), the 1S mass $m_t^{\rm 1S}$, and the PS masses 
		$m_t^{\rm PS}(\mu_f)$  for $\mu_f=20$, $50$ and $80$~GeV taking $\overline m(\overline m)=163$~GeV as input. Conversion errors are smaller than the dot sizes. Compiled from results given in Ref.~\cite{Hoang:2017btd}.}
\end{figure}

To define an adequate scale dependent short-distance mass for observables where the mass sentivity is affected by QCD dynamics with momentum scales $R<m_t$, one switches to an effective description in analogy to the well-known Foldy-Wouthuysen transformation~\cite{Foldy:1949wa}. Here the virtual off-shell and hard top quark quantum effects are also absorbed into the mass (or `integrated out'), but without absorbing the soft top quark dynamics. A mass scheme that realizes this concept is the {\bf MSR scheme} $m^{\rm MSR}(R)$~\cite{Hoang:2008yj,Hoang:2017suc,Hoang:2017btd} defined for $R< m_t$ by the relation
\begin{equation}
\label{eq:MSRdef}
m_t^{\rm MSR}(R) = m_t^0 \left\{1+\Big(\frac{\alpha_s(R)}{\pi}\Big)\left[\frac{1}{\epsilon} +\ln(4\pi e^{-\gamma_E})  + A^{\rm fin}(m_t^0/R)  \right]\right\}\, - \, R\Big(\frac{\alpha_s(R)}{\pi}\Big) A^{\rm fin}(1)\,+\, \ldots
\,.
\end{equation}
The MSR mass absorbs self-energy corrections coming from scales above $R$ (see again Fig.~\ref{fig:schemes1} and compare to the $\overline{\mbox{MS}}$ mass). The definition above makes is possible that $R$ can be chosen to be much smaller than $m_t$ consistent with the renormalization group. In practical applications $R$ should be set to the momentum scale that governs the soft quantum fluctuations affecting the mass sensitivity of the observable (e.g.\ the width of the resonance in the $m_t^{\rm reco}$ distribution). In Eq.~(\ref{eq:MSRdef}) the second peculiar looking term linear in $R$ is essential since in the difference 
$m_t^0 A^{\rm fin}(m_t^0/R)-R A^{\rm fin}(1)$ all linear sensitivity to soft momenta (in the a top resonance frame) cancels, so that the 'bad' soft momentum contributions in the top self-energy of Eq.~(\ref{eq:sigmaself}) are still left to cancel in calculations for processes as was the case for the $\overline{\mbox{MS}}$ mass $\overline m_t(\mu)$. The MSR mass is therefore also a short-distance mass.
The price to pay is that the MSR mass has a renormalization group equation {\it linear} in $R$,
\begin{equation}
\label{eq:MSRevolv}
\frac{\rm d}{{\rm d} \ln R} \,m_t^{\rm MSR}(R)\, = \, - \frac{4}{3}\,R \,\Big(\frac{\alpha_s(R)}{\pi}\Big)\,+\, \ldots\,,
\end{equation}
which is a generic requirement for a short-distance mass scheme with a renormalization scale $R<m_t$~\cite{Voloshin:1992wg,Bigi:1997fj}. The MSR mass is prototypical for the class of 'low-scale short-distance masses' devised in the last two decades for quantum field theory calculations for  $B$ mesons, heavy quarkonia and top resonance physics, such as the kinetic~\cite{Czarnecki:1997sz}, the 1S~\cite{Hoang:1998ng,Hoang:1998hm,Hoang:1999ye},
PS~\cite{Beneke:1998rk}, RS~\cite{Pineda:2001zq} and jet mass~\cite{Jain:2008gb}. The MSR mass is, however, the only low-scale short-distance mass that is, like the $\overline{\rm MS}$ mass, defined directly from the quark self-energy diagrams. For $R=\overline m_t(\overline m_t)$ it differs from the $\overline{\mbox{MS}}$ mass $\overline m_t(\overline m_t)$ only by corrections related to
two-loop self-energy corrections from virtual top quark loops. Therefore it can be considered as the natural extension of the  $\overline{\mbox{MS}}$ mass concept for renormalization scales below $m_t$, as was advocated in Ref.~\cite{Hoang:2017suc,Hoang:2017btd}. 
Interestingly, the MSR mass $m_t^{\rm MSR}(R)$ is also numerically close to the other low-scale short-distance masses at their respective intrinsic scales. See Fig.~\ref{fig:schemes2}, where $m_t^{\rm MSR}(R)$ is shown together with the 1S mass and PS masses at three representative scales. From a numerical point of view, the $\overline{\mbox{MS}}$ mass $\overline m(\mu)$ and the MSR mass $m_t^{\rm MSR}(R)$ can be related to each other and to other low-scale short-distance masses with a precision of $30$~MeV or better using available results from the literature, see e.g.\ Refs~\cite{Marquard:2015qpa,Hoang:2017suc}\footnote{Current uncertainties in the strong coupling do not allow to reach this precision in the relation of all short-distance masses. From Eqs.~(\ref{eq:poleMSR}) and (\ref{eq:poleMSbar}) one can see that an uncertainty in the strong coupling $\alpha_s(M_z)$ of $0.001$ results in a parametric uncertainty in the relation between a low-scale short-distance top mass and the $\overline{\rm MS}$ top mass of around $70$~MeV.}. 
This $30$~MeV precision represents the principal theoretical limitation of short-distance top mass determinations and is fully adequate for the era of hadron colliders.

The most frequently used top quark mass scheme in perturbative computations is the {\bf pole scheme} $m_t^{\rm pole}$, where self-energy corrections from all scales are absorbed into the mass,
\begin{equation}
\label{eq:poledef}
m_t^{\rm pole} = m_t^0 \left\{1+\Big(\frac{\alpha_s(\mu)}{\pi}\Big)\left[\frac{1}{\epsilon} +\ln(4\pi e^{-\gamma_E})  + A^{\rm fin}(m_t^0/\mu)  \right]\right\}\,+\, \ldots\,,
\end{equation}
see Fig.~\ref{fig:schemes1}.
By definition, in the pole scheme all perturbative quantum correction to the pole of the propagator vanish. Thus $m_t^{\rm pole}$ is the mass of the top quark states that appear in parton level scattering amplitudes
in the approximation where top quarks are treated as real external (or 'asymptotic') particles~\cite{Lehmann:1954rq,Lehmann:1957zz}. Because the mass of the formally defined top quark asymptotic states is renormalization scale invariant, infrared finite and gauge-invariant at the level of perturbation theory~\cite{Tarrach:1980up,Kronfeld:1998di}, it appears to be unique and physical, at least at parton level. However, as already mentioned above, due to the top quark's color charge, this concept is actually unphysical considering a precision of $0.5$~GeV or below. This can be seen from the fact that, due to the term $A^{\rm fin}$, the expression on the RHS of Eq.~(\ref{eq:poledef}) depends linearly on the way infrared momenta are regularized. (Recall the example of a gluon mass regulator shown in Eq.~(\ref{eq:Agluonmass}).) This means that the pole of the top quark propagator (and the meaning of the mass of a top quark asymptotic state) depends linearly on the way infrared momenta (in a top resonance frame) are treated. However, what is commonly called 'the pole mass $m_t^{\rm pole}$' in the context of QCD is defined {\it strictly} within dimensional regularization, where e.g.\ the IR regulator gluon mass term $\lambda$ shown in Eq.~(\ref{eq:Agluonmass}) does not arise.

The pole mass $m_t^{\rm pole}$ is obtained from the MSR mass $m_t^{\rm MSR}(R)$ taking the formal limit $R\to 0$, so that the MSR mass can be seen as a scheme that kind of interpolates between 
the pole and the $\overline{\mbox{MS}}$ masses. For finite $R$ the relation between the pole and MSR masses reads
\begin{equation}
\label{eq:poleMSR}
m_t^{\rm pole}-m_t^{\rm MSR}(R) \, = \, \frac{4}{3} \Big(\frac{\alpha_s(R)}{\pi}\Big)\,R\,+\, \ldots\,.
\end{equation}
In comparison, the relation between the pole and $\overline{\rm MS}$ masses has the form
\begin{equation}
\label{eq:poleMSbar}
m_t^{\rm pole}-\overline m_t(\mu) \, = \, \frac{4}{3} \Big(\frac{\alpha_s(\mu)}{\pi}\Big)\,\overline m_t(\mu)\,+\, \ldots\,.
\end{equation}
From the conceptional point of view, the MSR mass 
$m_t^{\rm MSR}(R)$ can also be seen as a scheme designed for observables where (virtual and real) QCD corrections below the scale $R$ are {\it unresolved} so that the self-energy corrections below $R$, which are not absorbed into $m_t^{\rm MSR}(R)$, are left to cancel with other (real or virtual) quantum fluctuations from scales below $R$ in a top resonance frame. In this context the pole mass is a scheme that is based on the unphysical view that virtual and real perturbative QCD corrections can be resolved down to arbitrarily small scales.

The unphysical character of the pole mass concept is reflected in the fact that the perturbative series for physical observables in the pole scheme carry the so-called {\bf `pole mass renormalon ambiguity'}~\cite{Bigi:1994em,Beneke:1994sw,Beneke:1998ui}. 
At this point let me briefly detour to explain some general aspects of renormalons, as they do not only arise for the pole mass. 	
Renormalons arise in {\it all} parton level cross sections $\hat\sigma$ computed in dimensional regularization, even for observables involving only massless quarks or gluons. This is because the perturbation series for all QCD observables are asymptotic series as already mentioned below Eq.~(\ref{eq:ABconversion}). This means that the terms in the perturbation series may decrease at low orders, but they
eventually adopt divergent patterns, which I call `turnover' in the following.
These divergent patterns in QCD perturbation theory were already discovered very early in the history of QCD~\cite{Gross:1974jv,Lautrup:1977hs,tHooft:1977xjm} and
can be caused by sensitivities to {\it physical} infrared QCD dynamical effects that are directly associated to specific types of nonperturbative corrections contained in $\sigma^{\rm NP}$. So, nonperturbative contributions to $\sigma^{\rm NP}$
having characteristic scaling behaviors in powers of $(\Lambda_{\rm QCD}/Q)^n$ are in one-to-one correspondence to specific types of
asymptotic divergent patterns~\cite{David:1983gz,Mueller:1984vh}. This correspondence is understood very well mathematically and the associated divergent patterns of the perturbative coefficients can be quantified analytically to all orders. The generic rule applies that the lower the power of $n$ is in a contribution to $\sigma^{\rm NP}$, the stronger is the associated asymptotic divergent pattern and the lower is the order of turnover. The formal mechanism that brings it all together is as follows:
When making predictions for the observable $\sigma^{\rm exp}$, the correction term $\sigma^{\rm NP}$ compensates, order-by-order in perturbation theory, the divergent patterns in $\hat\sigma$ {\it and at the same time} adds the corresponding  physical nonperturbative corrections. This connection is a fundamental aspect of QCD predictions of the form in Eq.~(\ref{eq:sigma}), where perturbative and nonperturbative contributions are separated and dimensional regularization is used to regulate infrared momenta~\cite{David:1983gz,Mueller:1984vh}.	In practice it may not be easy to realize this mechanism at high perturbative order, but for most applications the available perturbative corrections appear to be below the order of turnover.

When making parton level predictions in the pole mass scheme, there is a renormalon that arises from a divergent pattern coming from the virtual non-self-energy corrections that are soft (in a top resonance frame) and left uncancelled. I have already emphasized this cancellation issue several times above, and this is why. 
The divergent pattern of this pole mass renormalon has the same mathematical structure as those patterns related to contributions to $\sigma^{\rm NP}$ linear in $\Lambda_{\rm QCD}$, and this is in one-to-one correspondence to the linear infrared sensitivity of the term $A^{\rm fin}$ illustrated in Eq.~(\ref{eq:Agluonmass}). As such, the pole mass renormalon is rather strong and its numerical impact and even the turnover point can be visible and relevant already at the low orders accessible to perturbative calculations. So the pole mass renormalon looks very much like 
an uncertainty due to some missing physical nonperturbative information to be remedied by a contribution in $\sigma^{\rm NP}$ linear in $\Lambda_{\rm QCD}$ (i.e.\ proportional to $\Lambda_{\rm QCD}({\rm d}/{{\rm d}m_t})\hat{\sigma}$). 
However, this view is incorrect, since the pole mass renormalon pattern
is an artificial problem tied to the unphysical concept of a top quark particle pole and not related to a physical effect (that would be encoded in $\sigma^{\rm NP}$ in Eq.~(\ref{eq:sigma})~\cite{Bigi:1994em,Beneke:1994sw,Beneke:1998ui}. 
Using instead a short-distance mass at an appropriate renormalization scale, this renormalon is just gone. 
Insisting on using the pole mass, however, one must truncate at the order where the correction is minimal, i.e.\ around the order of turnover. The inability to do that in a unique way in practice (and even in principle) results in an ambiguity in the determination of $m_t^{\rm pole}$ and is called the `pole mass renormalon ambiguity".  
Interestingly, the ambiguity is unchanged even if the top quarks finite lifetime is accounted for, which shifts the top quark propagator pole to a complex $p^2$ value~\cite{Smith:1996xz}. This underlines the unphysical character of the pole mass renormalon. 

\begin{figure}[t]
	\includegraphics[scale=0.8]{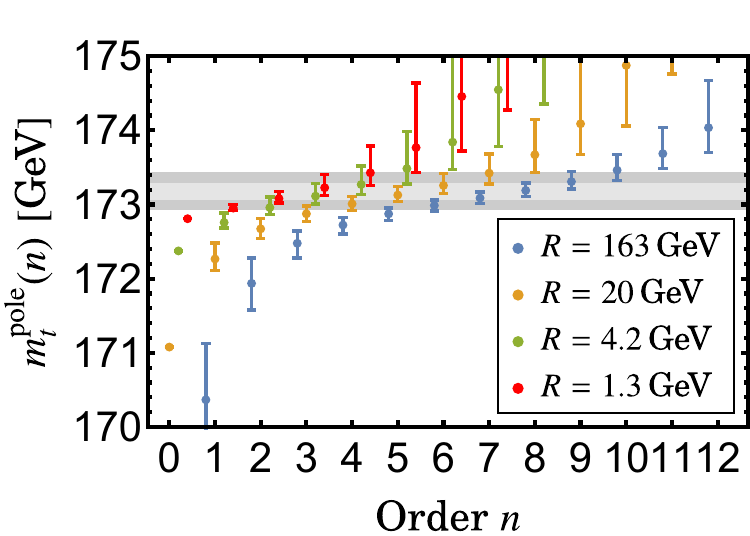}
	\caption{\label{fig:polemass} Pole mass determinations as a function of order $n$ from $m_t^{\rm MSR}(R)$ for $R=163$, $20$, $4.2$ and $1.3$~GeV using the series~(\ref{eq:poleMSR}), which is known to all orders. The values of $m_t^{\rm MSR}(R)$, shown as the dots at order $n=0$, are determined from $\overline m_t(\overline m_t)=163$~GeV using matching at $R=\overline m_t(\overline m_t)=163$~GeV and the R-evolution equation~(\ref{eq:MSRevolv}) at four loops and finite botton and charm masses with $\overline m_b(\overline m_b)=4.2$~GeV and $\overline m_c(\overline m_c)=1.3$~GeV. The error bars arise from renormalization scale variations. Compiled from Ref.~\cite{Hoang:2017btd}. The light gray band represents the pole mass including the renormalon ambiguity estimated in Ref.~\cite{Beneke:2016cbu} and the dark band the one obtained in Ref.~\cite{Hoang:2017btd}.  }
\end{figure}

Interestingly, the divergent pattern of the pole mass renormalon happens to grow so rapidly with order that for many quantities the explicitly calculated coefficients are already completely saturated by it~\cite{Hoang:2017btd,Hoang:2017suc,Beneke:2016cbu,Peset:2018ria}.\footnote{
The series for $m_t^{\rm pole}-m_t^{\rm MSR}(R)$ and $m_t^{\rm pole}-\overline m_t(\mu)$ in Eqs.~(\ref{eq:poleMSR}) and (\ref{eq:poleMSbar}), respectively, have been computed explicitly up to ${\cal O}(\alpha_s^4)$~\cite{Tarrach:1980up,Gray:1990yh,Chetyrkin:1999ys,Chetyrkin:1999qi,Melnikov:2000qh,Marquard:2007uj}.
The large order renormalon asymptotics of the pole mass renormalon has been shown to saturate the ${\cal O}(\alpha_s^3)$ and ${\cal O}(\alpha_s^4)$ coefficients and
the coefficients at ${\cal O}(\alpha_s^n)$ for $n>4$ are therefore known with a precision of few percent from their asymptotic behavior~\cite{Beneke:2016cbu,Hoang:2017suc}.
}  The ambiguity is also reflected in the form of Eq.~(\ref{eq:poleMSR}), where the limit $R\to 0$ can apparently only be taken by crossing the Landau pole in $\alpha_s(R)$. Since this would generate a nonperturbative contribution to the pole mass (which is unphysical), the limit $R\to 0$ is taken keeping the scale $\mu$ in $\alpha_s(\mu)$ finite. 
This is another way to understand how 
the divergence pattern in the coefficients of perturbative series in the pole mass scheme arises.
The pattern is illustrated in Fig.~\ref{fig:polemass}, where the pole mass is determined from the MSR mass $m_t^{\rm MSR}(R)$ at different values of $R$ from Eq.~(\ref{eq:poleMSR}) as a function of truncation order $n$. For the orders $n>4$ the asymptotic estimates are used, see footnote~5.
The observed pattern is representative of the behavior of pole mass determinations from physical observables (not affected linearly by other kinds of soft QCD effects) where the mass dependence is governed by QCD dynamics at the scale $R$. For $R\gsim m_t$ (relevant for total inclusive cross sections, see blue points and error bars) the order of turnover is 7 or 8, and one needs to go for many orders to get to the final range of $m_t^{\rm pole}$ values (gray bands). For $R\lsim 10$~GeV (relevant for some differential cross sections with kinematic top mass sensitivity, see red and green points and error bars) the order of turnover is 2 or 3. Here one can reach the final range of $m_t^{\rm pole}$ values already at orders accessible with available perturbative computations (see Sec.~\ref{sec:FO}) and even the tree-level results are close to it.\footnote{ 
This also illustrates that for perturbative computations for observables where $R$ is large, it is more natural to use $m_t^{\rm MSR}(R)$ (for $R<m_t$) or $\overline m_t(R)$ (for $R>m_t$) as mass schemes, while $m_t^{\rm pole}$ or $m_t^{\rm MSR}(R)$ are more natural for observables where $R$ is small.}
The size of the ambiguity and final range for $m_t^{\rm pole}$ (around the order of turnover) can be formally proven to be independent of $R$, and the size of the ambiguity can be formaly shown to be of order $\Lambda_{\rm QCD}$~\cite{Bigi:1994em,Beneke:1994sw,Beneke:1998ui}. Recent analyses quantified it as $110$~MeV~\cite{Beneke:2016cbu} and $250$~MeV~\cite{Hoang:2017btd} (width of the gray bands) using all available theoretical information\footnote{
	These estimates were obtained for 
	finite charm and bottom quark masses. For $m_c=m_b=0$ the ambiguity was estimated as $70$~MeV in Ref.~\cite{Beneke:2016cbu} and $180$~MeV in Ref.~\cite{Hoang:2017btd}.
	The dependence on the charm and bottom masses reflects the strong sensitivity of the pole mass on small momenta. It arises because the ambiguity is depending on the leading $\beta$-function coefficient of the strong coupling $\alpha_s$ which increases when the number of massless quarks is decreased. 
}. Figure~\ref{fig:polemass} also underlines that
when numerically converting between pole mass and short-distance mass values, it is essential to truncate at the order of turnover (related to the vertical location of the gray bands), which may differ from the perturbative order used in the calculation. I adopt the approach from Ref.~\cite{Hoang:2017btd} when quoting numerical values for the difference between $m_t^{\rm pole}$ and short-distance masses.

It should be stressed, however, that even when a short-distance mass scheme is used for a parton level top mass dependent cross section $\hat{\sigma}$, one still has to deal with possible contributions in $\sigma^{\rm NP}$, and in particular with those depending linearly on $\Lambda_{\rm QCD}$ which also
affect mass determinations in a linear way. Such linear nonperturbative effects are unavoidable for LHC observables because the hard processes generating the top mass sensitivity always involve top states (single top, $t\bar t$, \ldots) in a non-singlet color configuration or depend on jets. For these observables the description of the physically observable top mass dependence always involves nonperturbative color neutralization processes 	
which enter $\sigma^{\rm NP}$ linearly, as I already emphasized at the end of Sec.~\ref{sec:principle}. 
For top mass determinations these nonperturbative color neutralization processes must be understood separately and disentangled from the top mass dependence to ever reach the principle theoretical uncertainty limit of $30$~MeV for a short-distance mass determination mentioned above.\footnote{There is the possibility of a partial cancellation between the pole mass renormalon divergent pattern and the renormalon pattern related to a mass-unrelated physical linear nonperturbative effect due to a sign difference in the associated divergent patterns in the perturbative series. Examples for such observables were discussed e.g.\ in Refs.~\cite{Hoang:2018zrp,Fleming:2007xt,FerrarioRavasio:2018ubr}, and it was demonstrated in Refs.~\cite{Hoang:2018zrp,Fleming:2007xt} that the respective renormalon patterns arise from different dynamical modes located in separated and factorizable physical phase space regions. I disagree with the argument made in Ref.~\cite{FerrarioRavasio:2018ubr} that in such a case the principle precision of a pole mass determination is higher than for a short-distance mass determination.} The limit can be approached for $e^+e^-$ colliders (see footnote~3), but it is very difficult to do so for LHC observables. However, the size of the physical color neutralization corrections at the LHC is observable-dependent and can in some cases be controlled field theoretically using QCD factorization or be small, see Refs.~\cite{Hoang:2017kmk} and \cite{FerrarioRavasio:2018ubr}, respectively, for related studies.

\section{STATUS OF TOP MASS DETERMINATIONS AT THE LHC }
\label{sec:status}

In this section I discuss the status of state-of-the-art top mass determinations, focusing on the experimental methods and theoretical tools employed in the analyses. Since there are already a number of excellent reviews on the experimental aspects of the measurements in Refs.~\cite{Tanabashi:2018oca,Deliot:2010ey,Incandela:2009pf,Schilling:2012dx}, on projections for the HL-LHC (see Ref.~\cite{Azzi:2019yne,CMS:2013wfa} and references therein) and on the status of the employed MC tools (see Refs.~\cite{Tanabashi:2018oca,Corcella:2019tgt}),
I refrain from a detailed technical presentation and rather concentrate on the conceptual aspects critical for LHC top mass measurements.

\subsection{Fixed-Order Calculations}
\label{sec:FO}

State-of-the-art fixed-order parton level computations, i.e.\ perturbation series for $\hat{\sigma}$ as expansions in powers of $\alpha_s$, have reached a high level of sophistication and are primarily based on numerical methods. For the production of on-shell top quark pairs, QCD corrections at next-to-next-to leading order (NNLO)~\cite{Czakon:2016ckf} are available including the resummation of QCD next-to-next-to leading (NNLL) logarithms involving the ratio of the top quark transverse momentum $p_T$ and $m_t$~\cite{Czakon:2018nun} and also accounting for NLO electroweak corrections~\cite{Czakon:2017wor}. In the narrow-width approximation (NWA) for the top quark, NNLO QCD calculations for on-shell top quark production and top quark decay~\cite{Gao:2012ja} have been combined to allow for fully differential predictions~\cite{Gao:2017goi}. These results neglect finite-lifetime effects and do not account for the summation of large logarithms of the ratio $\Gamma_t/m_t$ related to the top quarks low-energy off-shell dynamics. Finite-lifetime effects have been included in fixed-order QCD NLO computations for the $W^+W^-b\bar b$ final state, including $W$ decays in leptons~\cite{Heinrich:2013qaa} or jets~\cite{Denner:2017kzu}. These calculations 
describe top production, top decay to $W^+W^-b\bar b$ final states and $W^+W^-b\bar b$ non-resonant production.
The latter results are less precise concerning QCD corrections and lack the resummation of logarithmic terms. NLO and higher fixed-order calculations provide reliable approximations with controlled top mass scheme dependence to $\hat\sigma$ for observables where the typical momenta of the QCD dynamics governing the mass dependence are of the size $m_t$ or larger. Relevant for top mass measurements, this includes the total inclusive $t\bar t$ cross section, the $t\bar t+$jet invariant mass $M_{t\bar t j}$ for values much larger than $2m_t$ and leptonic distributions away from kinematic threshold structures (kinks, shoulders, endpoints). Including the summation of logarithms of the ratio $p_T/m_t$ further provides reliable parton level descriptions of the $p_T$ and the $t\bar t$ invariant mass $M_{t\bar t}$ distributions in the boosted top region where $p_T \gg m_t$ or $M_{t\bar t}\gg 2 m_t$. Interestingly, almost all fixed-order calculations are available only in the pole mass scheme. Making parton level predictions in short-distance top mass schemes requires a reexpansion of the perturbative series using Eqs.~(\ref{eq:poleMSR}) or (\ref{eq:poleMSbar}) and the computation of mass derivatives. See Ref.~\cite{Langenfeld:2009wd} for a dedicated calculation of the total inclusive $t\bar t$ cross section at NNLO in the $\overline{\rm MS}$ top mass scheme.

\begin{figure}[t]
	\includegraphics[scale=0.4]{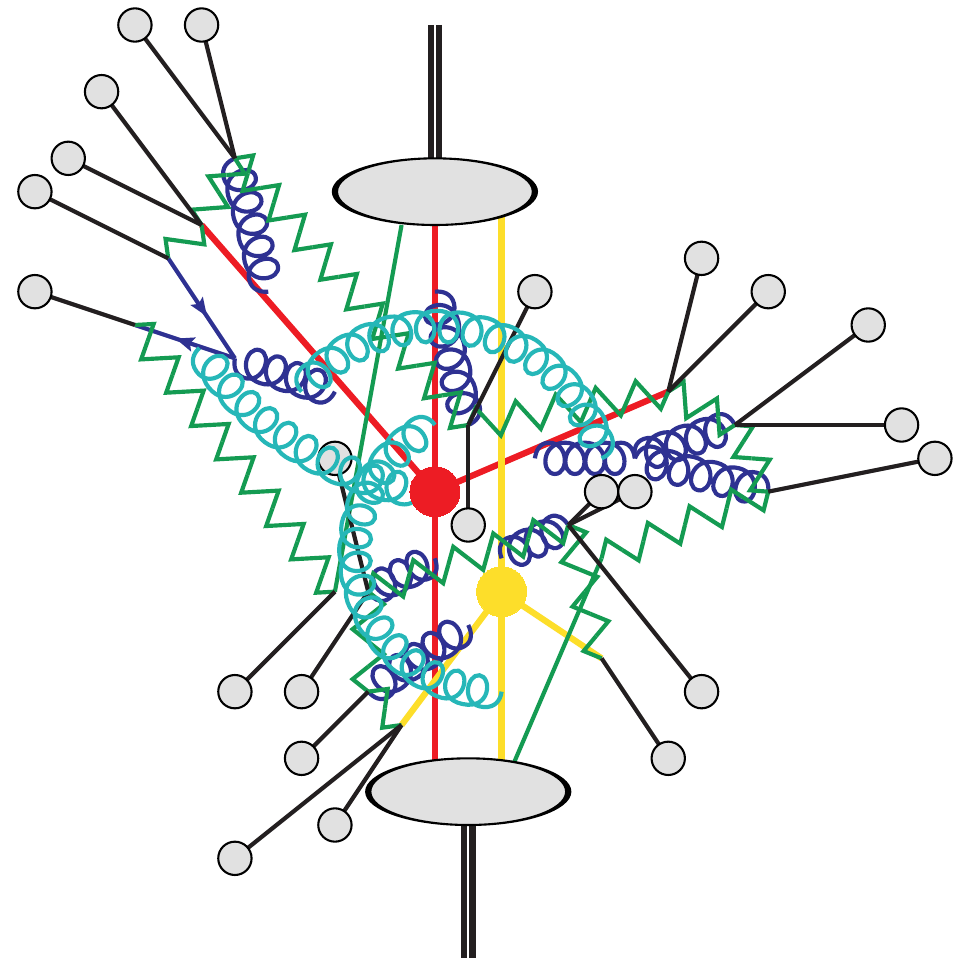}
	\caption{\label{fig:MC1} Generic picture for components of MC event generators.  }
\end{figure}

\subsection{Multipurpose MC Event Generators}
\label{sed:MMC}

Multipurpose Monte-Carlo event generators~\cite{Bellm:2015jjp,Sjostrand:2014zea,Gleisberg:2008ta} (MMCs) form the backbone of essentially 
all experimental analyses at the LHC. They are used to simulate all processes spanning from the colliding protons to the emergence of the observable hadrons. MMCs are used to 
design novel observables and measurements, for detector simulations, and to determine efficiencies and acceptances.
As illustrated in  Fig.~\ref{fig:MC1}, they combine the quark and gluon (parton) structure of the colliding protons (big gray blobs), tree-level leading order (LO) matrix elements for the hard parton interactions (red), a parton shower (PS) that describes the branching of the hard partons into lower energy partons (dark blue) and a hadronization model. The latter turns the high-multiplicity partonic states that emerge after the PS terminates into the observable hadronic particles, accounting for the color flow in the large-$N_c$ limit (small gray blobs for hadrons and green zigzag lines for color correlations). The hard matrix elements and the PS provide descriptions for $\hat\sigma$ in the collinear and soft limits, where fixed-order calculations are insufficient due to large logarithmic terms. These descriptions can be NLL precise for certain simple classes of observables such event-shapes, but are in general less precise even though they can still provide a description adequate for experimental simulations~\cite{Dasgupta:2018nvj,Cormier:2018tog}. State-of-the art PSs are either based on angular ordered 
coherent branching (CB)~\cite{Marchesini:1983bm}
(as used as the default in the {\scshape Herwig}~\cite{Bellm:2015jjp} MMC family) or on transverse momentum 
ordered dipole showering~\cite{Nagy:2005aa} (as used in the {\scshape Pythia}~\cite{Sjostrand:2014zea} and 
{\scshape Sherpa} MMCs~\cite{Gleisberg:2008ta} and optionally also in {\scshape Herwig}~\cite{Platzer:2009jq}). Differences between the two PS types arise for example in the treatment of non-global observables, where CB has intrinsic restrictions, or in momentum recoil 
effects, where dipole showering is based on a local treatment for each parton branching leading to precision issues for global observables. The description of the proton structure in terms of parton distribution function and the hadronization models provide approximate descriptions for $\sigma^{\rm NP}$.
The hadronization models are based on the concepts of decaying 
clusters~\cite{Webber:1983if} or the breaking of QCD strings~\cite{Andersson:1983ia}. Their parameters are not fixed from first principles QCD but through
the tuning procedure, i.e.\ by demanding that the MMCs reproduce a certain set of experimental differential reference cross sections. This allows the MC generators to provide adequate descriptions even when the PS description is less precise. 

\begin{figure}[b]
	\includegraphics[scale=0.5]{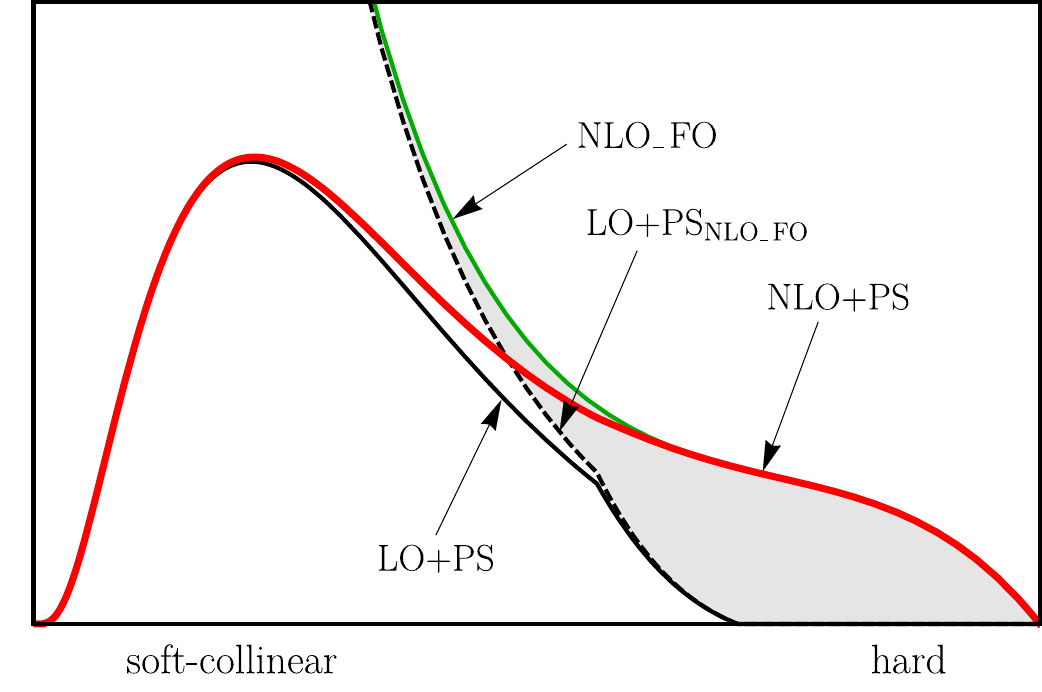}
	\caption{\label{fig:MC2} Generic structure of a kinematic distribution with a top mass sensitive kinematic threshold structure in the soft-collinear region (on the left side) obtained by NLO-matched MMCs (NLO+PS, red) and unmatched MMCs (LO+PS, solid black). The distribution at NLO in QCD fixed-order perturbation theory (NLO\textunderscore FO, solid green) is singular and diverges in the soft-collinear limit. The parton shower evolution of the unmatched MMCs sums the leading logarithmic singular terms to all orders in fixed-order perturbation theory leading to a physically meaningful approximation with Sudakov suppression in the soft-collinear limit (LO+PS, solid black). The matching procedure adds the difference between unmatched MMC description expanded out to NLO (LO+PS$_{\rm NLO\textunderscore FO}$, dashed black) and the NLO QCD fixed-order (NLO\textunderscore FO, solid green) results, both of which are singular, to the tail of the unmatched MMC distribution in the region dominated by hard radiation events (gray area, on the right hand side).  Since at the NLO fixed-order level the first hard emission arises from the first emission generated by the PS, this elevates the first hard PS emission of the NLO-matched MMCs (NLO+PS, red) to full NLO precision in QCD. Some distributions have a tail on both sides of the soft-collinear region. 
	}
\end{figure}

The precision of PSs in MMCs can be elevated by matching them with NLO matrix elements (referred to as NLO+PS). Matched generators such as MacGraph5\textunderscore aMC@NLO~\cite{Frixione:2002bd,Alwall:2014hca} or the POWHEG~\cite{Alioli:2010xd} procedure and
related methods available in {\scshape Herwig}~\cite{Platzer:2011bc} and  {\scshape Sherpa}~\cite{Hoeche:2011fd}
improve the description of the {\it first} hard PS emission to NLO precision (typically with transverse momenta larger than $10$~GeV) but leave the soft and collinear emissions and hadronization provided by the underlying MC generators unchanged. Figure~\ref{fig:MC2} illustrates generically how the matching affects the distribution of a top mass sensitive kinematic distribution, see the caption for a more detailed explanation.
MMCs share in an observable-dependent way the aspects of first-principle calculations as well as model-descriptions, where the primary goal is the good description for experimentally observable quantities. Obtaining reliable estimates of the theoretical uncertainties of the MMC descriptions is therefore a nontrivial task. 
There is an ongoing effort to improve the theoretical basis of MMCs and the methods to estimate their uncertainties for observable quantities, see e.g. Refs.~\cite{Bellm:2016rhh,Bendavid:2018nar,Dasgupta:2018nvj,Ravasio:2018lzi}.

For top quark physics, mostly the {\scshape Pythia}~\cite{Sjostrand:2014zea} and {\scshape Herwig}~\cite{Bellm:2015jjp,Bellm:2017bvx} event generators are employed. 
It is an essential aspect of all experimental top quark measurements to properly estimate the theoretical or model uncertainties of the MMC descriptions. The common approach of the experimental collaborations is to analyse the variations obtained from a few different MMC implementations that are considered reasonable. Limitations in state-of-the-art MMCs, particularly relevant for LHC top mass measurements, concern subtle issues such as color-reconnection~\cite{Gieseke:2012ft,Argyropoulos:2014zoa,Christiansen:2015yqa,Gieseke:2017clv,Gieseke:2018gff}, multiparticle interactions (yellow in Fig.~\ref{fig:MC1})~\cite{Sjostrand:2004ef,Bahr:2008dy}, the precise determination of parameters of the hadronization models~\cite{Corcella:2017rpt} or finite lifetime effects~\cite{Jezo:2016ujg,Heinrich:2017bqp}. In addition, MMCs used for state-of-the-art LHC analyses only contain LO matrix elements for the top decay. A serious principle limitation is that all massive quark PSs are theoretically based on the quasi-collinear (i.e.\ boosted top) approximation, while the bulk of the top mass measurements rely on top events with relatively low top quark transverse momenta $p_T\sim 100$~GeV and velocities $v_t\sim 0.5$. How this restriction affects current top mass measurements is to the best of my knowledge unknown and also not quantified for top mass measurements. 
Furthermore, the parton level descriptions of top mass sensitive kinematic threshold structures provided 
by NLO-matched MMC generators are not elevated to subleading QCD precision. This is because sharp threshold structures are governed by soft and collinear radiation and hadronization effects.
Examples of observables subject to this issue are all kinematic observables reconstructed from a single top quark such as its reconstructed invariant mass  $m_t^{\rm reco}$ or kinematic endpoint regions for variables such as the lepton energy $E_\ell$ or the lepton and b-jet invariant mass $M_{b\ell}$. Also the reconstructed $t\bar t$ invariant mass $M_{t\bar t}$ in the threshold region where $M_{t\bar t}\approx 2m_t$ is subject to this issue, due to soft radiation effects related to the $t\bar t$ pair produced with small relative velocity in a color-octet state, Coulomb binding corrections and coherence effects in the simultaneous weak decay of the $t\bar t$ pair. 
I reiterate that observables with such threshold structures are responsible for the high top mass sensitivity of the direct mass measurements. But it should in turn also be mentioned that NLO-matched MMCs can provide NLO reliable parton level approximations for observables  $\hat{\sigma}$  where the top mass sensitivity is generated exclusively by hard interactions (referred to as observables with indirect top mass sensitivity below Eq.~(\ref{eq:muevolv})). Examples are the total cross section or the mass variables $M_{t\bar t}$ and $M_{t\bar t j}$ far above threshold. So if the NLO-matching procedure uses the identification $m_t^{\rm MC} = m_t^{\rm pole}$, a measurement of $m_t^{\rm MC}$ from such hard interaction dominated observables can indeed be considered as a pole mass measurement.

In this context, measurements of the top quark mass are more subtle than measurements of physical observables (hadron and lepton momenta, lifetimes, hadronic and jet cross sections, \ldots). This is 
because the top mass and its couplings are not physical observables, but theoretically defined Lagrangian parameters. For their measurement the MMC employed has to provide perturbative ($\hat\sigma$) and nonperturbative descriptions ($\sigma^{\rm NP}$) {\it separately} consistent with QCD, such that the mass and coupling parameters of the generator retain a definite and observable independent relation to the QCD Lagrangian parameters. 
This relation is diluted or even lost to the extent that the tuning compensates for conceptual deficiencies of the PS regardless of whether the MMC describes the data well. This is particularly subtle for the top quark mass parameter $m_t^{\rm MC}$ since the MMC has to reliably simulate all color neutralization (linear in $\Lambda_{\rm QCD}$) and finite-lifetime (linear in $\Gamma_t$) effects consistent with the SM. 
This issue is the core of the MMC top mass interpretation problem related to the direct top mass measurements. Only limited quantitative knowledge on this complex set issues exists today.

\begin{figure}[t]
	\centering
	\begin{minipage}[b]{0.51\textwidth}
		\includegraphics[width=\textwidth]{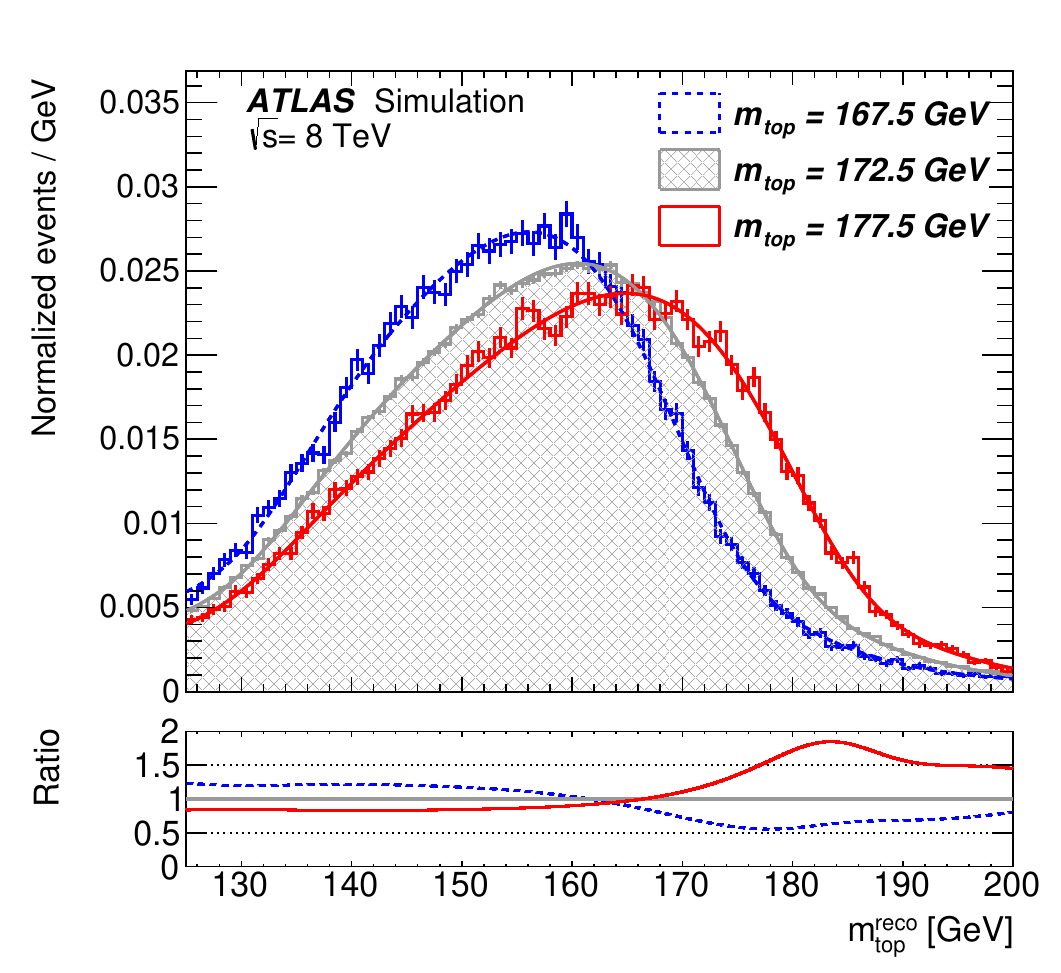}	
	\end{minipage}\hfill
	\begin{minipage}[b]{0.49\textwidth}
		\includegraphics[width=0.87\textwidth]{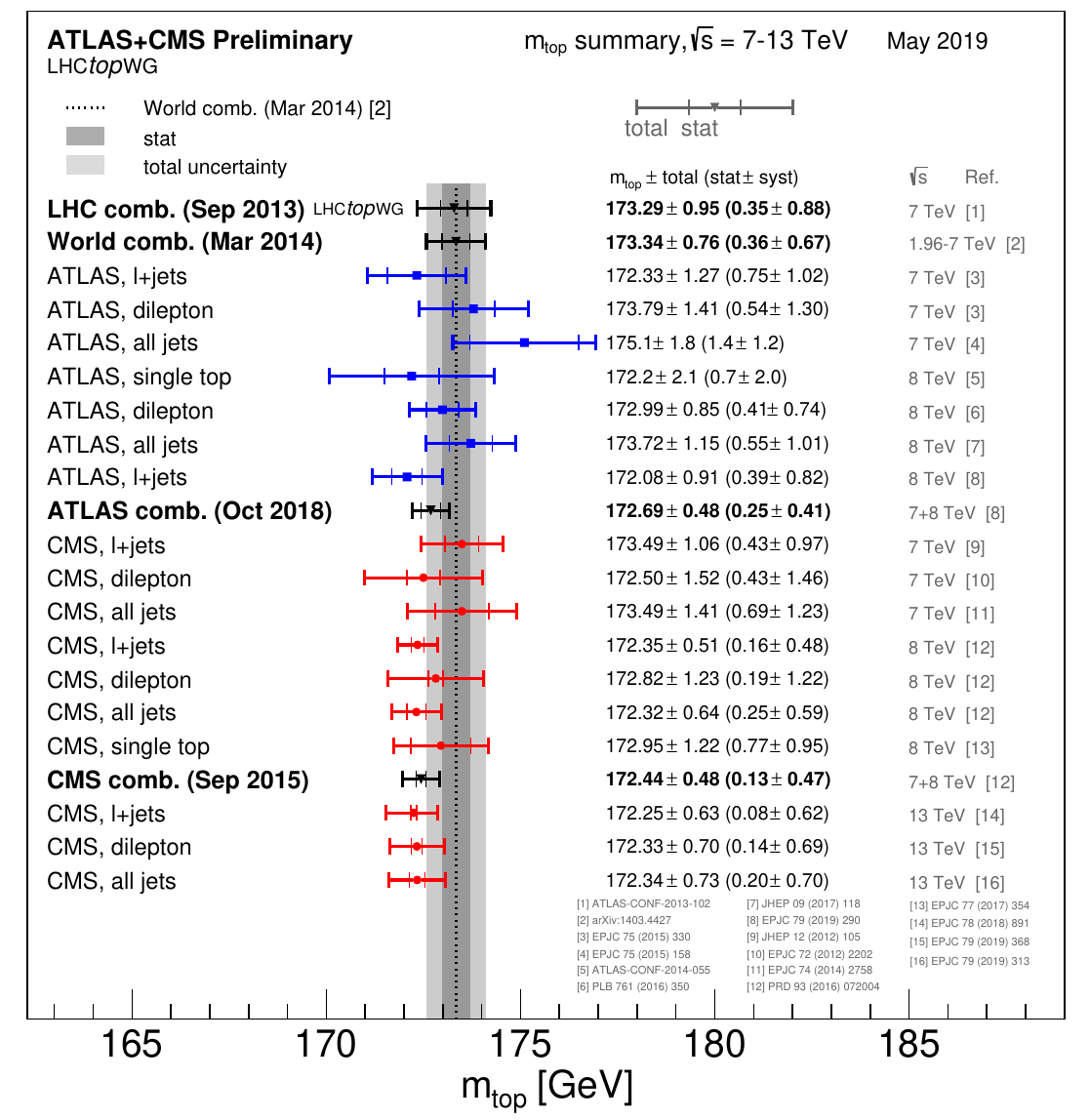}
	\end{minipage}
	\caption{\label{fig:ATLAStemplate} Left panel: Top mass dependence of the reconstructed top invariant mass $m_t^{\rm reco}$ obtained from top decays into three jets from MMC simulations in Ref.~\cite{Aaboud:2018zbu} (ATLAS collaboration). Right panel: Collection of recent LHC direct top mass measurements.}
\end{figure}

\subsection{Experimental analyses}

The {\bf direct measurements} are the most precise top quark mass extractions carried out at the LHC. 
They are based on kinematic observables constructed from reconstructed top decay products (light quark and b quark jets, leptons) for the different accessible top decay (semi-leptonic or hadronic) and top production modes ($t\bar t$ and single top events). For the template method the $b$-jet-lepton invariant mass $M_{\ell b}$ (dilepton $t\bar t$ and single top events) and reconstructed top invariant mass $m_t^{\rm reco}$ 
(see left panel of Fig.~\ref{fig:ATLAStemplate})
distributions are used. For the ideogram and matrix element methods the likelihood for a whole reconstructed final state to be compatible with a $t\bar t$ production hypothesis is determined event-by-event. Both approaches rely fully on the PS and hadronization components of MMCs for the theoretical description, so that it is the mass parameter $m_t^{\rm MC}$ which is extracted from the best fits or the highest cumulative likelihood. A summary of all state-of-the-art direct top mass measurements is shown in the right panel of Fig.~\ref{fig:ATLAStemplate}. The current world average quotes
$m_t^{\rm MC} = 172.9 \pm 0.4$\,GeV~\cite{Tanabashi:2018oca}. The latest CMS and ATLAS combinations have yielded 
$m_t^{\rm MC} = 172.26 \pm 0.61$\,GeV~\cite{Sirunyan:2018mlv} (see~\cite{Sirunyan:2017huu} for a measurement using single top events), and 
$m_t^{\rm MC} = 172.69 \pm 0.48$\,GeV~\cite{Aaboud:2018zbu}, respectively.
But I also want to recall the final Tevatron combination which obtained
$m_t^{\rm MC} = 174.34 \pm 0.64$\,GeV~\cite{Tevatron:2014cka}.\footnote{I believe that much could be learned from knowing the reasons for the discrepancy between the Tevatron and the LHC measurements. The impact a recalibration of the jet energy scale for the Tevatron D0 lepton+jet direct mass measurement~\cite{Abazov:2015spa} was analyzed in Ref.~\cite{Siikonen:2020yie}.}
As already mentioned, the $M_{\ell b}$ and $m_t^{\rm reco}$ variables employed for the template method are examples for observables whose MMC description is not improved by the NLO matching. The ideogramm and matrix element methods are based on observables of the same kind, because such observables have the highest mass sensitivity for the reconstructed decay products. Significant work is invested in the determination of the systematic uncertainties by the experimental collaborations. These efforts, however, do not provide insights concerning the interpretation problem of $m_t^{\rm MC}$, which -- as long as the issue is unresolved -- must be viewed as an additional systematic error in the relation of $m_t^{\rm MC}$ to a top mass scheme defined in field theory.

\begin{figure}[t]
	\centering
	\begin{minipage}[b]{0.5\textwidth}
		\includegraphics[width=\textwidth]{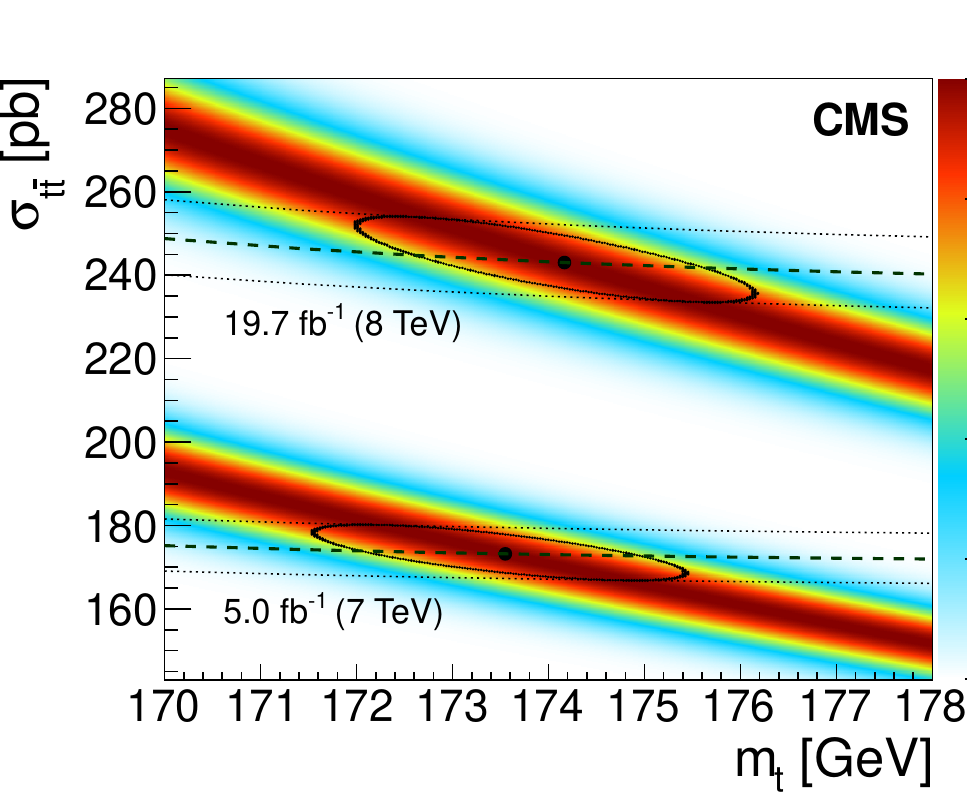}		
	\end{minipage}\hfill
	\begin{minipage}[b]{0.45\textwidth}
		\includegraphics[width=\textwidth]{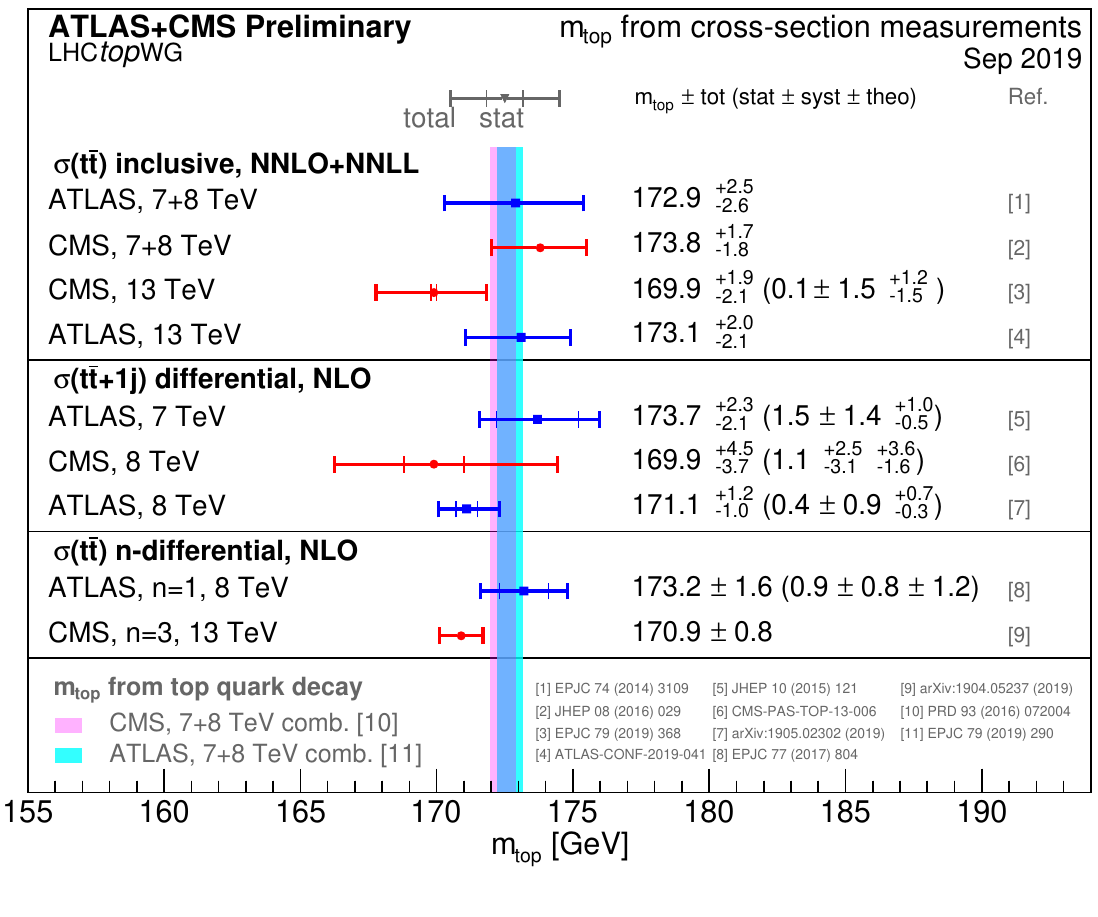}
	\end{minipage}
	\caption{\label{fig:LHCmeasurementspole} Left panel: Top mass dependence of the measured (black lines) and theoretical predicted (dark shaded band) inclusive $t\bar t$ cross section and the resulting best $m_t^{\rm pole}$ fit range, from Ref.~\cite{Khachatryan:2016mqs} (CMS collaboration)  based on LHC $7$ and $8$~TeV data.
		Right panel: Summary of recent pole mass measurements.}
\end{figure}

So-called {\bf pole mass measurements} are based on the inclusive and differential $t\bar t$ cross sections, for which theoretical parton level predictions expressed in the pole mass scheme from NNLO+NNLL calculations for the total cross section $\sigma(t\bar t+X)$~\cite{Czakon:2013goa} or NLO-matched MC generators for the reconstructed $t\bar t+$jet invariant mass $M_{t\bar tj}$~\cite{Alioli:2013mxa}, (di)leptonic variables~\cite{Frixione:2014ala} and $t\bar t$ invariant mass $M_{t\bar t}$ are available. A summary of these measurements is shown in the right panel of Fig.\ \ref{fig:LHCmeasurementspole}, and the current world average quotes
$m_t^{\rm pole} = 173.1 \pm 0.9$\,GeV~\cite{Tanabashi:2018oca}. The inclusive $t\bar t$ cross section and the invariant masses $M_{t\bar t}$ and $M_{t\bar tj}$ (away from the lower threshold at $2m_t$) are examples of observables where the top mass sensitivity is indirect, i.e.\ exclusively tied to hard interactions. For them, parton level predictions at NLO (or higher) and NLO-matched MC generators carry NLO information on the mass scheme. Furthermore, for these observables the resolution scale $R$ for the QCD dynamics governing the mass sensitivity (see Fig.\ \ref{fig:polemass}) is of order or larger than $m_t$. One can therefore expect that the theoretical errors of the parton level prediction may be further reduced when even higher order fixed-order or resummed calculations become available or when the $\overline{\mbox{MS}}$ top mass scheme is employed.
Inclusive cross section measurements yielded 
$m^{\rm pole}_t= 172.9^{+2.5}_{-2.6}$\,GeV (ATLAS, $7$ and $8$~TeV data)~\cite{Aad:2014kva},
$m^{\rm pole}_t= 173.8^{+1.7}_{-1.8}$\,GeV (CMS, $7$ and $8$~TeV data)~\cite{Khachatryan:2016mqs} and
$m^{\rm pole}_t= 169.9^{+2.0}_{-2.2}$\,GeV (CMS, $13$~TeV data)~\cite{Sirunyan:2018goh}.\footnote{The analysis of Ref.~\cite{Sirunyan:2018goh} also studied the strong correlation between the extracted top mass, the value of the strong coupling $\alpha_s(M_Z)$ and the employed set of parton distributions functions~\cite{Alekhin:2017kpj,Dulat:2015mca,Harland-Lang:2014zoa,Ball:2017nwa}.
The quoted lower value for $m^{\rm pole}$ is based on
a set of parton distribution functions~\cite{Alekhin:2017kpj} that is determined in a simultaneous fit with $\alpha_s$. The associated range of $\alpha_s$ values is below that of the world average.  
The analysis also determined the $\overline{\mbox{MS}}$ top mass $\overline m_t(\overline m_t)$ based on the calculations of Ref.~\cite{Langenfeld:2009wd}.} The relatively larger errors in comparison to the direct measurements result from the uncertainty in the normalization of the inclusive cross section (dominated by gluon luminosity uncertainties and renormalization scale variation in the cross section fixed-order calculations) and its relatively weaker dependence on $m_t$, see the left panel of Fig.~\ref{fig:LHCmeasurementspole}. 
A recent $M_{t\bar tj}$ measurement by the ATLAS collaboration yielded $m^{\rm pole}_t= 171.1^{+1.2}_{-1.1}$\,GeV~\cite{Aad:2019mkw}, which is more precise since the distribution exhibits a mass sensitive broad hump. 
A measurement using leptonic distributions by the ATLAS collaboration obtained 
$m^{\rm pole}_t= 173.2\pm 1.6$\,GeV~\cite{Aaboud:2017ujq}. It should be pointed out that for leptonic distributions the color neutralization effects I mentioned earlier indirectly affect the momentum of the decaying $W$ boson and cannot be avoided.  
A CMS analysis including the total inclusive cross section, the $M_{t\bar t}$ and the top pair rapidity $y_{t\bar t}$ distributions and a simultaneous $\alpha_s$ and gluon distribution fit obtained $m^{\rm pole}_t= 170.5\pm 0.8$\,GeV~\cite{Sirunyan:2019zvx} and poses some tension with the pole mass world average mentioned above. For the latter analysis I would like to remark that the smaller error compared to the inclusive cross section measurements above partly results from the inclusion of the $M_{t\bar t}$ distribution which is kinematically sensitive to the top mass in the threshold region $M_{t\bar t}\approx 2m_t$. 
This is an issue to be examined thoroughly for achieving reliable theoretical descriptions, because the theoretical fixed-order calculations employed for the analysis to determine the top mass are based on the approximation where $M_{t\bar t}$ is defined from the 4-momenta of on-shell top quarks. On the other hand, NLO-matched MMC descriptions, used to relate the reconstructed observable $M_{t\bar t}$ distribution to the theory calculation, do not have subleading QCD precision for $M_{t\bar t}$ in the threshold region. Furthermore, a large fraction of the $t\bar t$ pairs is produced in color-octet configurations, for which the effects of soft QCD radiation are significant.

A number of {\bf alternative methods} to measure $m_t$ have been proposed, which are based on differential cross sections with respect to alternative mass sensitive variables constructed from top decay products. 
The observables include the $M_{T2}$ variable and variants of it~\cite{Lester:1999tx,Chatrchyan:2013boa},
the shape of b-jet and B meson energy distributions~\cite{Agashe:2016bok},
the $J/\psi$ and lepton invariant mass~\cite{Khachatryan:2016pek,Khachatryan:2016wqo}, secondary vertices from $b$ quark fragmentation~\cite{Khachatryan:2016wqo}. They are also motivated having in mind the kinematics of a decaying top particle. These observables are affected by issues similar to the direct measurements albeit with differing systematics and leading to larger uncertainties. They can also be seen as $m_t^{\rm MC}$ measurements and are consistent with the direct measurements. Using the fact that the sensitivity to soft and nonperturbative dynamics can be reduced by jet grooming techniques~\cite{Butterworth:2008iy,Ellis:2009me,Krohn:2009th,Larkoski:2014wba}, it was suggested to use the mass of a fat and groomed boosted top quark jet, for which factorized QCD predictions with field theoretical control of the top mass scheme and nonperturbative effects can be determined~\cite{Hoang:2017kmk}. In Ref.~\cite{Kawabata:2016aya} the $\gamma\gamma$ invariant mass spectrum $M_{\gamma\gamma}$ was suggested as a top mass sensitive variable since it shows a glitch due to large QCD phases and Coulomb bound state effects when
$M_{\gamma\gamma}\approx 2m_t$. Predictions of the $\gamma\gamma$ mass observable in principle allow to control the top mass scheme systematically, but I remark again that LHC produces significant amounts of $t\bar t$ pairs in a color-octet state. Due to the effects of radiation that is soft in $t\bar t$ c.m.\ as well as the lab frame, precise and reliable predictions of $M_{\gamma\gamma}$ are therefore significantly more involved than for the analogoue $t\bar t$ threshold cross section in $e^+e^-$ annihilation~\cite{Hoang:2013uda,Beneke:2015kwa}, and are still to be achieved. Furthermore, the $\gamma\gamma$ mass method requires HL-LHC to be competitive with the current pole mass measurements uncertainties.

Overall, current direct and pole mass measurements show good mutual agreement, but the discriminating power of the pole mass measurements is somewhat lower. One can expect that the theoretical uncertainties of pole mass measurements may be further reduced when the corresponding next higher order perturbative calculations or improved theoretical approaches become available. An additional reduction of theoretical uncertainties may be achieved when, instead of the pole mass scheme, appropriate scale-dependent short-distance mass schemes such as $\overline{\mbox{MS}}$ or MSR are employed. This should, however, also be accompanied with some substantially increased understanding concerning a number of systematic effects influencing the size and shape of the related differential cross sections which currently affect these top mass measurements at the level of $1$~GeV or larger. It requires dedicated work for the pole mass measurements to approach the numerical precision of the direct measurements quoted by the experimental collaboration. However, one has to keep in mind that the direct measurements suffer from an additional uncertainty related to the $m_t^{\rm MC}$ interpretation problem.

\section{THE CONTROVERSY}
\label{sec:controversy}

The controversy described in Sec.~\ref{sec:intro} is about the question of whether or not the interpretation problem of $m_t^{\rm MC}$ is large compared to the experimental uncertainties quoted for the direct top mass measurements.
The arguments for the two viewpoints can be paraphrased in a concrete form as follows. 

There has been the view, advocated in Refs.~\cite{Nason:2016tiy,Nason:2017cxd}, to write  
\begin{equation}
\label{eq:option1}
m_t^{\rm MC}=m_t^{\rm pole} + \Delta_{m_t}^{\rm MC}
\end{equation}
where it is assumed that the identification of the MMC top mass parameter with the pole mass is appropriate to very good approximation
and the term $\Delta_{m_t}^{\rm MC}$ is related to the approximate MMC theory description and modeling. The term $\Delta_{m_t}^{\rm MC}$ is an uncertainty {\it in addition} to the uncertainties quoted by the experimental collaborations, but it is argued to be much smaller than these such that it is appropriate to identify $m_t^{\rm MC}=m_t^{\rm pole}$. This view is based on the following argumentation: First, the parton level components of MMCs (hard matrix elements, parton shower (PS)) are good approximations to perturbative computations made in the pole scheme. Second, the PS can be assumed to provide a good approximation to soft and collinear perturbative radiation at in principle all soft scales for the observables entering the direct measurements. Since self-energy corrections (which are virtual) do not show up explicitly in the PS, they are effectively absorbed into the quark mass parameter $m_t^{\rm MC}$ which would result in its identification with the pole mass. The pole mass renormalon ambiguity is argued to be relevant in the sense that the identification of $m_t^{\rm MC}$ with the pole mass only breaks down when experimental uncertainties approach the size of the pole mass ambiguity (which means that $m_t^{\rm MC}-m_t^{\rm pole}$ is limited in size by the ambiguity). This, however, using e.g.\ the estimate of $110$~MeV for the pole mass ambiguity from Ref.~\cite{Beneke:2016cbu}, does not happen for the current top mass measurements and the projections for the HL-LHC. 

The other view, advocated in Refs.~\cite{Hoang:2008xm,Hoang:2014oea}, is to write
\begin{equation}
\label{eq:option2}
m_t^{\rm MC,Q_0}=m_t^{\rm MSR}(R_0) + \Delta_{m_t}^{\rm MC}(R_0,Q_0)
\end{equation}
where $R_0$ is an appropriate scale. The argumentation is as follows: In state-of-the-art MMCs the PS evolution terminates at a scale $Q_0$ around $1$~GeV (called the `shower cut') which keeps the strong coupling governing the PS in the perturbative regime and avoids that the number of partons becomes too large and computationally unmanageable.
Since the PS in MMCs is an approximation to perturbative soft and collinear radiation for scales above $Q_0$, {\it all} (real and virtual) radiation at scales below $Q_0$ is treated as unresolved and thus left to combine and cancel. Therefore the self-energy corrections from scales below $Q_0$ are not absorbed into $m_t^{\rm MC}$. This implies that the generator mass depends on the shower cut $Q_0$ (and in principle also the type of the PS), is close to the MSR mass $m_t^{\rm MSR}(R_0)$ for $R_0\propto Q_0$ and thus a short-distance mass like the MSR mass. The relation can be computed if $Q_0$ is treated as a factorization scale such that the PS is only used in the perturbative regime and not to model nonperturbative effects (as it should be for a first principles perturbative calculation). So $\Delta_{m_t}^{\rm MC}(R_0\propto Q_0,Q_0)$ is a finite perturbatively computable term scaling like $\alpha_s(Q_0)\times Q_0\sim 0.5$~GeV. It is not captured by the uncertainties quoted by the experimental collaborations and may not be smaller than these. To determine it reliably, detailed additional insights into the perturbative precision and structure of PSs and the physical meaning of their shower cut $Q_0$ are mandatory. This also implies a level of scrutiny on the theoretical precision of PSs and hadronization models
in state-of-the-art MMCs beyond of what is presently imposed, to find out whether $\Delta_{m_t}^{\rm MC}(R_0,Q_0)$ is observable independent or has a nonperturbative contribution.  
The size of the pole mass renormalon ambiguity plays no role in Eq.~(\ref{eq:option2}) which is a relation between two short-distance masses. 

The second view is conceptually more involved than the first. The controversy thus boils down to different judgment on (a) whether the first view on the smallness of $\Delta_{m_t}^{\rm MC}$ in Eq.~(\ref{eq:option1}) indeed applies or whether the formulation of
Eq.~(\ref{eq:option2}) is required\footnote{An explicit computation gives $m_t^{\rm pole}-m_t^{\rm MSR}(R_0=1.3\,\mbox{GeV})=360\pm 250$~MeV for finite charm and bottom masses~\cite{Hoang:2017btd}, see Fig.~\ref{fig:polemass}.}  
and (b) whether the impact of the shower cut on the perturbative components of the MMCs is negligible so that $Q_0$ is merely a parameter of the hadronization model or whether $Q_0$ is an infrared factorization scale at the interface between the perturbative and nonpertubative components of the MMCs that can (and must) be quantified analytically.
It should be also stressed that even though there is a controversy,  given
that $Q_0$ is a relatively small scale of around $1$~GeV, we are talking about differences and effects at the level of $0.5$ to $1$~GeV, but not more than that. In the context of QCD, worries that $m_t^{\rm MC}$ may be close to the $\overline{\mbox{MS}}$ mass $\overline m_t(\overline m_t)$ (which would constitute differences at the level of $10$~GeV) are unfounded. Furthermore, there is overall agreement in the demand that MMCs need to be improved to gain a higher level of precision concerning the quality of their PSs and hadronization models to reduce systematic uncertainties. For the second view this is a necessary condition to determine $\Delta_{m_t}^{\rm MC}(R_0,Q_0)$ from first principles analyses, but clearly
all methods to determine the top mass would benefit from such improvements.

\section{RECENT DEVELOPMENTS}
\label{sec:recent}

\subsection{Numerical Size of the Interpretation Problem}
\label{sec:numerical}

While initially only qualitative arguments for the interpretation problem for $m_t^{\rm MC}$
were available~\cite{Hoang:2008xm,Hoang:2014oea,Moch:2014tta,Nason:2016tiy,Nason:2017cxd}, 
recently some quantitative studies appeared, which have shed some light on the numerical aspect of the issue. 
In Ref.~\cite{Kieseler:2015jzh} a combined analysis using the direct method and a pole mass
measurement using the inclusive cross section was carried out yielding that $m_t^{\rm pole}-m_t^{\rm MC}$ is not larger than $2$~GeV.
In Ref.~\cite{Butenschoen:2016lpz} the numerical relations $m_t^{\rm MC}=m_t^{\rm pole}+(0.57\pm 0.29)$~GeV and 
$m_t^{\rm MC}=m_t^{\rm MSR}(1\,\mbox{GeV})+(0.18\pm 0.23)$~GeV were
obtained from fitting a NNLL+NLO calculation for the 2-jettiness distribution in the 
resonance region for boosted top production in 
$e^+e^-$ annihilation~\cite{Fleming:2007xt} to {\scshape Pythia}~8.2~\cite{Sjostrand:2014zea} pseudo-data samples. The result of this calibration study included a rigorous estimate of
nonperturbative uncertainties of the analytic NNLL+NLO calculation, since hadronization corrections can be rigorously desribed by a factorized shape function~\cite{Fleming:2007qr}.
An analogous analysis for the LHC was performed in Ref.~\cite{Hoang:2017kmk} 
using soft-drop groomed~\cite{Larkoski:2014wba} jet mass distributions at NLL+LO, obtaining
compatible but less precise results.

The results of the analyses in Refs.~\cite{Kieseler:2015jzh,Butenschoen:2016lpz} are consistent with the view that the interpretation problem for $m_t^{\rm MC}$
is limited to the level of $0.5$ or $1$~GeV. In Ref.~\cite{Butenschoen:2016lpz}
also valuable quantitative results concerning the two views of Eqs.~(\ref{eq:option1}) and (\ref{eq:option2}) were provided.
In particular, the term $\Delta_{m_t}^{\rm MC}$ in Eq.~(\ref{eq:option1}) has been shown to be about $0.5$~GeV (i.e.\ of the same size as the uncertainties quoted for the LHC direct top mass measurements) for an $e^+e^-$ process where {\scshape Pythia} (and all major MMCs) can be trusted to perform with a much higher precision. It is therefore conservative to conclude that
the error in identifying $m_t^{\rm MC}$ with the pole mass is at least of the same size as the direct measurement uncertainties quoted by the experimental collaborations. Furthermore, the small difference between $m_t^{\rm MC}$ and $m_t^{\rm MSR}(1\,\mbox{GeV})$ supports the second view that $m_t^{\rm MC}$ and MSR mass at a low scale are closely related. This motivates to study $\Delta_{m_t}^{\rm MC}(R_0,Q_0)$ in Eq.~(\ref{eq:option2}) from first principles and to invest work generalizing the $e^+e^-$ result for boosted top quark toward top production at the LHC. 

In Ref.~\cite{Hoang:2018zrp} such a first principles study was initiated, still for boosted top quark production in $e^+e^-$ annihilation. I review the results of this study in the following section.

\subsection{First Conceptual Insights}
\label{sec:conceptual}

To start the discussion, let us write down the relation between the MMC top mass parameter and the pole mass as
\begin{equation}
\label{eqn:mMCmpole}
m_t^{{\rm MC},Q_0} \, = \, m_t^{\rm pole} \,+\, \Delta^{\rm pert}_{\rm MC}(Q_0) \, + \, \Delta^{\rm non-pert}_{\rm MC}(Q_0) \, + \,\Delta^{\rm MC}\,. 
\end{equation} 
It presents a generalized unbiased version of Eqs.~(\ref{eq:option1}) and (\ref{eq:option2})  that makes the potential shower cut dependence of the MMC top mass parameter explicit and serves to visualize the issues that need to be understood. As written down, none of the three $\Delta$ terms on the RHS is accessible via the error estimates carried out by the experimental collaborations. The three $\Delta$ terms may even have different signs. Furthermore, all quantities except the pole mass $m_t^{\rm pole}$ are in principle MMC-dependent, which is indicated by the sub- and superscripts `MC'.  
The term $\Delta^{\rm pert}_{\rm MC}(Q_0)$ represents perturbative corrections starting at ${\cal O}(\alpha_s)$ related to the (kind of) PS and the PS cutoff used by the MMC. 
The sum $m_t^{\rm pole} \,+\, \Delta^{\rm pert}_{\rm MC}(Q_0)$ can be written in any top mass scheme, so for the intended conceptual discussion it does not matter which scheme we pick.
Let's use the pole mass since it is mostly used for fixed-order calculations. 
The term $\Delta^{\rm non-pert}_{\rm MC}(Q_0)$ stands for possible effects coming from the MMC hadronization model affecting the meaning of $m_t^{{\rm MC},Q_0}$ and should not be confused with the nonperturbative corrections the hadronization model generates in the description of physical observables.  It carries an argument $Q_0$ since it may depend on the PS set up, if the PS does not carry a definite precision consistent with QCD.
The term $\Delta^{\rm MC}$ encodes shifts due to MMC related systematic uncertainties which are physically unrelated to the dynamics of the top quark per se, but may indirectly affect the theoretical meaning top mass parameter. Effects contributing to $\Delta^{\rm MC}$ may include e.g.\ effects from color reconnection or the $b$-jet modelling and may be observable dependent. It would be evidence that $\Delta^{\rm non-pert}_{\rm MC}(Q_0)$ and $\Delta^{\rm MC}$ are sizeable for state-of-the-art MMCs if different measurements of $m_t^{\rm MC}$ are inconsistent. In a perfect MMC that made parton level calculations consistent with QCD to subleading order and had hadronization models behaving fully consistent with QCD for all processes, $\Delta^{\rm non-pert}_{\rm MC}(Q_0)$ as well as $\Delta^{\rm MC}$ would be negligibly small, and $m_t^{{\rm MC},Q_0}$ would be observable-independent. We could then simply calculate $\Delta^{\rm pert}_{\rm MC}(Q_0)$ from an analytic solution of the PS algorithm (with finite $Q_0$) for a simple mass sensitive observable and a comparison with the corresponding partonic QCD calculation.
In the analysis of Ref.~\cite{Butenschoen:2016lpz} already mentioned above, the sum of the three $\Delta$ terms was quantified as $(0.57\pm 0.28)$~GeV for the {\scshape Pythia}~8.2 MMC and the  $e^+e^-$ 2-jettiness distribution, but no information on the size and interplay of 
$\Delta^{\rm pert}_{\rm MC}(Q_0)$, $\Delta^{\rm non-pert}_{\rm MC}(Q_0)$ and $\Delta^{\rm MC}$ was acquired.
Such a differentiated knowledge is, however, mandatory to allow for first principles conclusions on the field theoretic interpretation of the MC top mass $m_t^{\rm MC}$. This is because sizeable contributions from $\Delta^{\rm non-pert}_{\rm MC}(Q_0)$ and $\Delta^{\rm MC}$ can make the meaning of $m_t^{{\rm MC},Q_0}$ observable-dependent and non-universal.

In Ref.~\cite{Hoang:2018zrp} a first principles study of Eq.~(\ref{eqn:mMCmpole}) was initiated by a dedicated analysis of 
the perturbative contribution $\Delta^{\rm pert}_{\rm MC}(Q_0)$. It was based on a combined analytical and numerical examination 
of the CB formalism for massive quarks~\cite{Gieseke:2003rz} that is the theoretical basis of the angular-ordered PS used in the {\scshape Herwig}~7 MMC.
The analysis was restricted in several ways: 
\begin{enumerate}
\item[(i)] The observable considered was the 2-jettiness event-shape distribution for boosted top pair 
production in $e^+e^-$ annihilation in the resonance region, which is a global observable and equivalent to the distribution of the sum of the squared hemisphere masses with respect to the thrust axis. For this observable the available  NNLL+NLO QCD computation~\cite{Fleming:2007xt,Butenschoen:2016lpz} is based on a factorization of {\it large-angle soft radiation} (i.e.\ radiation that is soft in the $t\bar t$ c.m.\ frame) and {\it ultracollinear radiation} (i.e.\ radiation that is soft in the respective resonance frames of the boosted top quarks)~\cite{Fleming:2007qr}. The results can therefore be immediately generalized to all $e^+e^-$ massive quark event shape type observables for which the ultracollinear dynamics is universal, but not to those employed for LHC top mass measurements. 
\item[(ii)] The use of an $e^+e^-$ event-shape variable such as 2-jettiness represents another physical restriction because the distribution is only sensitive to QCD radiation in the production stage of the top quarks while the effect of final-state radiation (off the top decay products) is power-suppressed. 
\item[(iii)] The NWA was employed, which does not account for finite-lifetime effects.
\item[(iv)] The angular ordered CB shower formalism was considered, which differs from the transverse momentum ordered dipole shower formalism.
\end{enumerate}

\noindent
In this context the following statments were proven by first principles computations and analytic as well as numerical studies:
\begin{enumerate}
	\item[{\bf 1.}] The consistent resummation of logarithms at NLL order in the singular resonance region, which carries   the kinematic mass-sensitivity, is mandatory and sufficient 
	to control the top mass scheme with NLO (${\cal O}(\alpha_s)$) precision. 
	\item[{\bf 2.}] The CB formalism for massive quarks~\cite{Gieseke:2003rz}, and thus also the angular ordered PS in {\scshape Herwig}~7, is NLL precise in the top mass sensitive singular resonance region and is fully equivalent with the NLL$^\prime$ QCD factorization predictions of Ref.~\cite{Fleming:2007xt,Butenschoen:2016lpz}.
	\item[{\bf 3.}] For vanishing infrared regularization (i.e.\ $Q_0=0$) the quark mass parameter appearing in the CB formalism at NLL (defined in an expansion in powers of $\alpha_s$ and logarithms) agrees with the pole mass $m_t^{\rm pole}$ to NLO, i.e.\ ${\cal O}(\alpha_s)$. This does not, however, apply to angular-ordered PSs because their evolution requires a finite shower cut $Q_0>\Lambda_{\rm QCD}$ to avoid infinite parton multiplicies and the strong coupling Landau pole. 
	\item[{\bf 4.}] In angular ordered PSs the shower cut $Q_0$ represents the minimal transverse momentum $p_{\perp}$ of radiated gluons or other partons that emerge from the showering and splittings. It can be also seen as a resolution scale or an infrared cutoff. An analysis of large-angle soft radiation as well as ultracollinear radiation with respect to effects linear in $Q_0$ was carried out for the PS and the QCD calculation. This amounts to using a finite $Q_0$ for the PS and imposing the $Q_0$ cut in the QCD calculation in the pole 
	scheme accounting also for the mass counterterm (which is absent in the CB algorithm\footnote{This issue is subtle and a potential source of misinterpretation.}).
	For the large-angle soft radiation the linear $Q_0$ cutoff dependence is physical and represents a factorization scale at the interface to a non-perturbative effects (which is known as the linear power correction $\alpha_0$~\cite{Davison:2008vx} or $\Omega_1$~\cite{Abbate:2010xh} in the tail of $e^+e^-$ event shape distributions). A change in $Q_0$ must therefore be compensated by a corresponding modification of non-perturbative contributions and does effectively not lead to a change at hadron level. For the ultracollinear radiation, terms linear in $Q_0$ are generated as well, but in the full QCD calculation their cumulative effect 
	in a smeared distribution or a moment cancels so that there is no physical net effect. However, the finite $Q_0$ value entails that all virtual (self-energy and non-self-energy) ultracollinear radiation effects become unresolved and cancel\footnote{I refer to the cancellation of linear ultracollinear quantum corrections already mentioned below Eq.~(\ref{eq:muevolv}) and before Eq.~(\ref{eq:MSRevolv}).} so that the pole of the top propagator is {\it shifted away} from $m_t^{\rm pole}$ (defined in the usual way without any infrared cut -- recall the discussion after Eq.~(\ref{eq:poledef})) by a term linear in $Q_0$. This shifted mass of the top propagator pole has been called {\it coherent branching} (CB) mass and reads
	\begin{equation}
	\label{eq:conclusions1}
	m_t^{\rm CB}(Q_0) \, = \,  m_t^{\rm pole} - \frac{2}{3}\,\alpha_s(Q_0)\,Q_0 \,+\, \ldots\,.
	\end{equation}
	The existence of a term linear in $Q_0$ on the RHS of this equality has precisely the same origin as the linear gluon mass term already shown in Eq.~(\ref{eq:Agluonmass}).
	Since the CB algorithm does not generate any self-energy corrections, the generator mass for finite $Q_0$ and at parton level is equal to $m_t^{\rm CB}(Q_0)$ rather than $m_t^{\rm pole}$, which
	implies $\Delta^{\rm pert}_{\rm CB,Herwig}(Q_0)= - \frac{2}{3}\alpha_s(Q_0)Q_0$.
	For boosted top quarks the effects linear in $Q_0$ on the large-angle soft and the ultracollinear radiation have an opposite sign, but also a different dependence with respect to the c.m.\ energy $Q$. Thus they can be analytically and numerically disentangled unambiguously at parton level. These linear contributions correctly exponentiate so that the mass change is consistently implemented in the resummed tower of logarithms.
	\item[{\bf 5.}] The CB mass $m_t^{\rm CB}(Q_0)$ is a short-distance mass, so its relation to other short-distance masses is not affected by the pole mass renormalon ambiguity. For example, the numerical relation between the CB and the MSR mass reads $m_t^{\rm MSR}(Q_0)-m_t^{\rm CB}(Q_0)=120\pm 70$~MeV for the 
	{\scshape Herwig}~7 shower cut $Q_0=1.25$~GeV, where $70$~MeV is an estimate of the missing two-loop correction. This allows to relate the CB mass to all other known short-distance masses with the same precision. It can also be seen that perturbation theory still works well at a scale of $1.25$~GeV (which is close to the charm quark mass).
	Reducing this perturbative uncertainty would require the determination of the ${\cal O}(\alpha_s^2)$ term in Eq.~(\ref{eq:conclusions1}) in the context of a NNLL order precise CB algorithm.
	The difference of the CB and the pole mass mass can be determined using the relation between the MSR and the pole mass shown in footnote~11. This gives $m_t^{\rm pole}-m_t^{\rm CB}(Q_0)=480\pm 260$~MeV which can be considered an all order relation that cannot ever be made more precise.  
\end{enumerate}

What can we learn from the results of the analysis? Let me start with some comments concerning its restrictions. 
The restriction to boosted top quarks goes hand in hand with the fact that both the CB formalism for massive quarks~\cite{Gieseke:2003rz} (as well as dipole-type shower algorithms) and the QCD factorization approach of Ref.~\cite{Fleming:2007xt} only apply in the quasi-collinear regime. 
For slow top quarks a QCD factorization approach disentangling the individual top quarks from each other does not exist and the use of branching algorithms is an extrapolation (even though no serious problems seemingly appear in the description of top event provided by MMCs). 
Conceptual and analytic first principles studies of the top quark generator mass for the bulk top quarks are therefore strictly speaking impossible with the current set of theoretical tools, and one has to rely on extrapolation studies starting from the boosted regime. On the other hand, this makes precision studies and top mass measurements with boosted top quarks, which become available with high statistics at the HL-LHC, interesting, see Ref.~\cite{
Sirunyan:2017yar,Aaboud:2018eqg,Sirunyan:2019rfa} for recent CMS and ATLAS measurements. 
The restriction to a global $e^+e^-$ dijet event-shape in the resonance region (where the top decay is treated fully inclusively) entails, as already mentioned, that the analysis is only sensitive to QCD radiation in the production stage of the top quarks.  Corresponding global event-shape observables at the LHC are considerably more involved due to the effects of initial state radiation, underlying event contamination and long-distance color correlations. For boosted top quarks, however, the basic simplicity of QCD factorized predictions for $e^+e^-$ collisions can be largely maintained also in hadron-hadron collisions if soft-drop groomed jet mass observables are considered~\cite{Hoang:2017kmk,Hoang:2019ceu}, so that an LHC study in analogy to Ref.~\cite{Hoang:2018zrp} is not unfeasible.
Furthermore,  $e^+e^-$ dijet event-shapes differ conceptually from the observables employed in the direct measurements which use observables differential in the top decay. This restriction can be lifted by considering more differential observables, and technology to do so is available from the vast knowledge obtained in the theory of B meson decays~\cite{Neubert:1993mb,Manohar:2000dt} and contemporary progress in factorized calculations~\cite{Becher:2014oda}. The soft-drop groomed top jet mass analysis in Ref.~\cite{Hoang:2017kmk} goes in that direction as well, but considers an observable not yet analyzed by the experimental collaborations. The restriction to the NWA has been applied since state-of-the-art PSs do not provide a systematic treatment of the top quark width. The {\scshape Herwig}~7 generator uses the NWA and {\scshape Pythia} is based on a NWA supplemented by throwing a random top mass value around the nominal top generator mass 
following a Breit-Wigner-type distribution. In the analysis~\cite{Heinrich:2017bqp} is was shown that, using different  approaches to treat the top decay and finite lifetime effects, can affect a top mass determination at the $0.5$ or even $1$~GeV level, so this restriction is a very serious one too. 
For the 2-jettiness QCD calculation the treatment of the leading finite width effects is well-understood~\cite{Fleming:2007qr}. 
Finally, the restriction to the CB formalism was motivated as it is designed to work well for global observables and allows for a straightforward analytic solution and comparison to the predictions of QCD factorization. This restriction can in principle be lifted by a dedicated study of the dipole shower formalism, which is more elaborate analytically.\footnote{In Ref.~\cite{Baumeister:2020mpm} a pure numerical analysis of the $Q_0$ dependence of the  $e^+e^-$ 2-jettiness distribution was carried out using the dipole-type parton shower implementation of Ref.~\cite{Dinsdale:2007mf}. The numerical results where found to be consistent with the statements described in point~4 above.}

It is clear that the restrictions just mentioned need to be lifted to resolve the interpretation problem for contemporary direct top quark mass measurements, but they reflect at the same time the principle limitations  of the state-of-the-art MMCs which should be remedied. 
Furthermore, definite knowledge on the nonperturbative terms $\Delta^{\rm non-pert}_{\rm MC}(Q_0)$ and $\Delta^{\rm MC}$ needs to be gained. One way to do so, is an analysis of the physical aspects of the hadronization models used in MMC from the perspective of observables for which definite statements on the first principles QCD structure of hadronization corrections are available. Each of the restrictions as well as of the nonperturbative terms, may have a numerical impact at the level of a few hundred MeV to $0.5$~GeV.
The results obtained in Ref.~\cite{Hoang:2018zrp} are therefore only a first step. The numerical analysis of 
Eq.~(\ref{eq:conclusions1}) demonstrates that the partonic contribution in Eq.~(\ref{eqn:mMCmpole}) is already of the size of the uncertainties quoted for current LHC direct mass measurements and that detailed analyses of all the terms on the RHS of Eq.~(\ref{eqn:mMCmpole}) is mandatory.
To the extent that the infrared-behavior of PS algorithms for ultracollinear radiation is universal and NLL precise, results of the kind of Eq~(\ref{eq:conclusions1}), which applies to {\scshape Herwig}~7, should be rather observable-independent and even apply to other MMCs, even though more studies are needed to substantiate this view. 

The minimal aspect to be learned from the analysis~\cite{Hoang:2018zrp} is that the identification of the direct mass measurements with the pole mass is field theoretically incorrect. There is clear evidence that the additional error associated with making the identification is at least of the same size as the quoted experimental direct measurement uncertainties. Furthermore, due to the different structure of the evolution variable of different PS algorithms, it appears natural that the physical meaning of the top mass parameters in different PSs should not be assumed to be universal. Overall, the analysis affirms that higher developed and more precise MMC (with respect to NLL accurate PSs and finite lifetime effects) are a necessary requirement to resolve the top mass interpretation problem.

\section{SUMMARY AND RECOMMENDATION}
\label{sec:conclusion}

In this review, I have presented an overview of the problems involved in the question how to interpret the direct top mass measurements, which quote the Monte-Carlo top mass parameter $m_t^{\rm MC}$, from a physical and conceptual perspective. They touch perturbative (parton showers (PSs) and finite width effects) as well as nonperturbative aspects and limitations of  multipurpose Monte-Carlo event generators (MMCs) and each may amount to effects at the level of several hundred MeV to half a GeV. 
Many of them go beyond the reach of the standard approaches used in high-energy collider physics today and require some novel avenues beyond the current paradigm of achieving higher theoretical precision by using MMCs matched to fixed-order perturbative computations.  
The top mass interpretation problem expresses the demand that, in order to measure theoretically defined QCD parameters at hadron colliders, MMCs themselves provide first principles QCD predictions which are accurate to subleading order in QCD in order to control the renormalization scheme of their QCD parameters. This is not the case for state-of-the-art MMCs.

For the observables used in the direct measurement, for which the top mass sensitivity is tied to kinematic threshold structures, this means that the PS algorithms should have NLL precision and that hadronization models are employed that implement nonperturbative effects consistent with QCD and the electroweak theory. For the top quark mass the radiation that is soft in a top quark resonance frame plays the most important role. It is probably unrealistic to ask for this level of precision for all observables. But for a number of key observables leading to high-precision top mass measurements with control over the mass scheme at NLO, I believe, such an achievement is realistic in the near future.
The relation of the MMC top mass parameter to any mass scheme and the question of universality and observable independence could then be obtained from computations rather than speculations. Developments in the direction of NLL precise PSs are already under way, e.g.\ concerning a more precise description of the parton splitting~\cite{Li:2016yez,Hoche:2017iem,Dulat:2018vuy}, the restriction of dipole-type showers for global observables~\cite{Dasgupta:2020fwr,Forshaw:2020wrq}, finite life-time effects~\cite{Brooks:2019xso} or full color coherence~\cite{Martinez:2018ffw,Forshaw:2019ver}, but there is still a long way to go.

How should one deal with the top mass interpretation problem today? It is well understood that the $m_t^{\rm MC}$ parameter obtained from direct top mass measurements is numerically close (i.e.\ within $0.5$ or maybe $1$~GeV) to mass schemes that are compatible with the top Breit-Wigner resonance. This includes the pole mass $m_t^{\rm pole}$ or the MSR mass $m_t^{\rm MSR}(R)$ for scales $R$ in the range of $1$ to $2$~GeV (i.e.\ close to the PS cutoff in MMCs and the top quark width $\Gamma_t$). This excludes the $\overline{\rm MS}$ mass $\overline m_t(\overline m_t)$ in QCD. In the analysis of Ref.~\cite{Hoang:2018zrp} the parton level relations between the top mass parameter $m_t^{\rm CB}(Q_0)$ in the angular-ordered PS of {\scshape Herwig}~7 and the pole and MSR masses were computed analytically
for an $e^+e^-$ observable, where the PS was shown to have NLL precision. The relation $m_t^{\rm CB}(Q_0)-m_t^{\rm pole}=- \frac{2}{3}\,\alpha_s(Q_0)\,Q_0$ was derived and is was shown that $m_t^{\rm CB}(Q_0)$ is a short-distance mass.  
Numerically, the results give
$m_t^{\rm pole}-m_t^{\rm CB}(Q_0)=480\pm 260$~MeV and $m_t^{\rm MSR}(Q_0)-m_t^{\rm CB}(Q_0)=120\pm 70$~MeV,
where $Q_0=1.25$~GeV is the {\scshape Herwig}~7 shower cutoff. 
In the work of Ref.~\cite{Butenschoen:2016lpz} the observable employed in Ref.~\cite{Hoang:2018zrp} was used to determine the corresponding relations numerically at hadron level by a calibration fit using a NNLL+NLO QCD calculation yielding $m_t^{\rm MC}-m_t^{\rm pole}= 570\pm 290)$~MeV and 
$m_t^{\rm MC}-m_t^{\rm MSR}(1\,\mbox{GeV})=0.180\pm 230$~MeV for {\scshape Pythia}~8.2. 
These results show that also the nonperturbative aspects of the interpretation problem is relevant for the state-of-the-art direct top mass measurements which have reached a precision of $0.5$~GeV. Clearly a deeper understanding is crucial to obtain a reliable and systematic high-precision top mass measurement at the HL-LHC. Much work is still needed to analyze how much the dynamical effects of the hadronization models affect the meaning of $m_t^{\rm MC}$ and to carry out similar analyses for observables closer to those used in the LHC measurements.

At this time there is no general consensus how to quantify the interpretation problem of the direct top mass measurments for making mass dependent theoretical predictions. It is left to the decision of the individual how to deal with the issue. I hope that this review provides the reader a deeper insight for her or his choice. 
Most often the identification $m_t^{\rm MC}=m_t^{\rm pole}$ is made, sometimes supplemented by adding another uncertainty of the order of the quoted experimental uncertainty. If this approach is adopted, I recommend, as a practical (neither very conservative nor very optimistic) attitude for the time being, that the uncertainty to be added is $0.5$~GeV accounting for the interpretation problem plus $250$~MeV for the pole mass renormalon ambiguity. 
As an {\it additional} option, which accounts for the existing evidence that the MMC generator top masses are short-distance masses and reflects a somewhat less conservative attitude, I recommend to use the identification 
$m_t^{\rm MC}=m_t^{\rm MSR}(1.3\,\mbox{GeV})$ adding an uncertainty of $0.5$~GeV quantifying the interpretation problem. In this approach the pole mass renormalon ambiguity is coming back in the conversion to $m_t^{\rm pole}$ for predictions made in the pole mass scheme (the outcome just differs by a $350$~MeV shift in the central value w.r.\ to the first approach). But the pole mass renormalon ambiguity is avoided completely when considering only predictions made in short-distance mass schemes. In this context one should employ the method explained in Sec.~\ref{sec:schemes} to convert between pole and short-distance masses.

\section{ACKNOWLEDGEMENTS}

I thank my collaborators Mathias Butenschoen, Bahman Dehnadi, Sean Fleming, Ambar Jain, Christopher Lepenik, Sonny Mantry, Vicent Mateu, Aditya Pathak, Simon Pl\"atzer, Moritz Preisser, Daniel Samitz, Ignazio Scimemi, Maxi Stahlhofen and Iain Stewart 
for their valuable contributions to work important to this review, I thank Katerina Lipka, Marcel Vos and the editor for comments and suggestions to the manuscript.
I acknowledge partial support by the FWF Austrian Science Fund under the Doctoral
Program (``Particles and Interactions'') No.\ W1252-N27, and the Projects No.~P28535-N27 as well as No.~P32383-N27
and by the European Union under the
COST action (``Unraveling new physics at the LHC through the precision
frontier'') No.\ CA16201. I thank Simon Pl\"atzer for providing Figs.~\ref{fig:MC1} and \ref{fig:MC2} and
Christopher Lepenik for providing all other figures.

\begin{appendix}

\section{GLOSSARY}
\label{sec:AppA}

\begin{tabular}{ll}
MMC: &  multipurpose Monte-Carlo event generator\\ 
SM: & Standard Model of elementary particle physics\\ 
QCD: & quantum chromodynamics\\
PS: & parton shower \\ 
c.m.: & center-of-mass\\
$Q_0$: & low-energy cutoff scale of the parton shower evolution\\ 
CB: & coherent branching \\ 
LHC: & Large Hadron Collider\\ 
HL-LHC & high-luminosity phase of the LHC \\
UV:&  ultra-violet (referring to large momenta or energies)  \\ 
$\Lambda_{\rm QCD}$: & scheme-dependent nonperturbative hadronization scale, typically around $200$~MeV \\
NWA:  & narrow width approximation\\ 
MSR: & subtraction renormalization scheme including power corrections down to the soft scale R\\ 
$\overline{\mbox{MS}}$: & conventional minimal subtraction scheme within dimensional regularization\\
LO:  & leading (lowest) order approximation in fixed-order perturbation theory \\
NLO:  & next-to-leading order approximation in fixed-order perturbation theory \\
NNLO:  & next-to-next-to-leading order approximation in fixed-order perturbation theory \\
NLL:&  next-to-leading order approximation in logarithm-resummed perturbation theory \\ 
NNLL:&  next-to-next-to-leading order approximation in logarithm-resummed perturbation theory \\ 
NLO+PS: & NLO-matched parton shower 
\end{tabular}

\end{appendix}


\end{document}